\documentclass[
 aps,prb,superscriptaddress,
 amsmath,amssymb,
reprint,%
]{revtex4-2}
\usepackage{CJK}

\usepackage{graphicx}% Include figure files
\usepackage{dcolumn}% Align table columns on decimal point
\usepackage{bm}% bold math
\usepackage[utf8]{inputenc}
\usepackage[T1]{fontenc}
\usepackage{subfigure}
\usepackage[T1]{fontenc}
\usepackage{txfonts}

\usepackage{color}
\usepackage{hyperref}
\usepackage[toc,page,titletoc]{appendix}
\usepackage[capitalise,nameinlink]{cleveref}

\usepackage[rightcaption]{sidecap}
\usepackage{setspace}

\usepackage{braket}

\newcounter{line}[part]

\newcounter{algo}[part]
\newcommand{\scaldot}{\boldsymbol{\cdot}}
\newcommand{\nrm}[1]{||\, #1 \,||}
\newcommand{\Tr}{\mathrm{tr}}
\newcommand{\ddif}{\mathrm{d}}
\newcommand\relphantom[1]{\mathrel{\phantom{#1}}}
\begin{document}

\title[]{State Initialization of a Hot Spin Qubit in a Double Quantum Dot by
         Measurement-Based Quantum Feedback Control}
%         Closed-Loop Measurement-Based Quantum Control}

\author{A.~Aarab}
\author{R.~Azouit}
\author{V.~Reiher}
\author{Y.~Bérubé-Lauzière}\email{Corresponding author: Yves.Berube-Lauziere@USherbrooke.ca} % Commande \email toute suite après l'auteur qui est le correspondant identifie le correspondant à l'auteur.
\affiliation{Institut quantique and Département de génie électrique et de génie informatique, Faculté de génie, Université de Sherbrooke, Sherbrooke, Québec, J1K 2R1, Canada}

\begin{abstract}
A measurement-based quantum feedback protocol is developed for spin state initialization in a gate-defined double quantum dot spin qubit coupled to a superconducting cavity. The protocol improves qubit state initialization as it is able to robustly prepare the spin in shorter time and reach a higher fidelity, which can be pre-set. Being able to pre-set the fidelity aimed at is a highly desired feature enabling qubit initialization to be more deterministic. The protocol developed herein is also effective at high temperatures, which is critical for the current efforts towards scaling up the number of qubits in quantum computers.
%In addition, it promises to help solve the problem of decoherence by stabilizing qubit states.
\end{abstract}

\maketitle

\section{Introduction}

%\vspace{-10pt}
% ANCIEN TEXTE
%Spin qubits are a promising technology for quantum computing owing to their long coherence times and spatial compactness. In order to control and manipulate the qubit state, it is mandatory to use non-destructive measurements. These measurements will be performed using a quantum dot coupled to a superconducting microcavity. To execute a quantum algorithm, the qubits need to be initialized to a known quantum state, tupically state~$\ket{0}$ associated with the qubit's ground state. The leading methods for initialization  use either relaxation~\cite{tuorila2017efficient} to the ground state, the successive projective~\cite{riste2012initialization}\cite{govia2015unitary} measurement or the feedback-based on a single shot measurement~\cite{ofek2016extending}\cite{Andersen-Molmer_ClosingQFB_PRA_2016}. Most of these methods, the fidelity does not exceed 99.8\% and the operating temperature should be in the range of $0~$K. In this work, we investigate a new approach for initializing a spin qubit in a double quantum dot (DQD) coupled to a superconducting cavity. It relies on a measurement-based quantum feedback protocol using weak dispersive measurements of the qubit's state via the cavity. We hereby present the design of a control system to drive the qubit towards the desired state in a gradual and continuous manner. This approach is also cheaper to implement experimentelly because the qubit can be operated in the range of temperatures of $2~$K with the fidelity more than 99.9\%.

Since Loss and DiVicenzo's initial proposal of a quantum computing architecture relying on spin states in coupled single-electron quantum dots~\cite{loss1998quantum}, steady efforts have been devoted to developing electron spin qubits in research laboratories~\cite{hanson2007publisher,zwanenburg2013silicon,benito2017input,Vandersypen-Eriksson_QCompSemiconSpins_2019,Noiri-Scappucci-Tarucha_FastUnivQGateThresh_2022}
and in industry with major players like Intel~\cite{Intel_IEEE_IEDM_2018,Zwerver_QuTech-Intel_QubitsAdvSemiConManuf_arXiv2021} and IBM~\cite{Kuhlmann_etal_AmbipolarQDsFinFET_2018,Geyer_Basel-IBM_2021}.
The long coherence times
%achievable with
of
electron spin qubits on the order of seconds in silicon~\cite{tyryshkin2012electron,zwanenburg2013silicon,hanson2007publisher,veldhorst2014addressable}, and their spatial compactness, along with the exquisite fabrication capabilities of the silicon electronics industry, make them great candidates for physical qubit implementations.

To execute quantum algorithms, a key requirement for a qubit according to the DiVicenzo criteria~\cite{divincenzo2000physical} is the ability to initialize it to a known quantum state, typically corresponding to the qubit's ground state and denoted $|0\rangle$.
In practice, it is necessary that qubit initialization be robust, \textit{i.e.} performed with high fidelity. This typically requires initialization to be carried out on a short timescale relative to the qubit's decoherence time. Protocols have been developed with this aim of reducing the initialization time and improving fidelity.
 
We herein develop and numerically evaluate a new measurement-based quantum feedback (MBQFB) approach for actively initializing a spin qubit with high fidelity in an architecture consisting of a Si/SiGe gate-defined double quantum dot (DQD) coupled to a microwave superconducting cavity ($\mu$WSCc), which is reminiscent of circuit quantum electrodynamics (cQED). This DQD-$\mu$WSCc architecture proposed in~\cite{beaudoin2016coupling,benito2017input,mi2018coherent} is a promising candidate for the fabrication of two-dimensional arrays of qubits with the long-range spin-spin connectivity required for quantum information processing.
%and error correction.
Such connectivity, which remains a great challenge~\cite{chanrion2020charge,mortemousque2018coherent,hendrickx2021four}, is achieved in this architecture through microwave photons in cavities to mediate the long-range spin-spin interactions, as has been demonstrated for superconducting qubits~\cite{blais2004cavity,wallraff2004strong,dicarlo2009demonstration}.
Using the large electric dipole moment of the electron charge state in a DQD through spin-charge hybridization, coherent interactions between single spins and microwave photons have already been demonstrated theoretically~\cite{benito2017input} and experimentally~\cite{viennot2015coherent,mi2018coherent}.

Several schemes for initializing qubits have been studied in cQED (see~\cite{tuorila2017efficient} and references therein).
They can be initialized to the ground state via passive thermal relaxation (which has also been used for electron spin qubits in quantum dots~\cite{Hanson_etal_IEDM2004,hanson2007publisher}),
% successive projective measurement~\cite{govia2015unitary},
or some form of feedback conditioned on the outcome of a single-shot measurement~\cite{riste2012initialization,Johnson-Siddiqi_etal_HeraldStatePrep_2012,Andersen-Molmer_ClosingQFB_PRA_2016}. These methods lead to limited fidelities and require very low operating temperatures in the tens to hundreds of~mK.

Our approach to spin qubit active initialization using MBQFB is made possible by the DQD-$\mu$WSCc architecture which allows weak dispersive and
quantum non-demolition (QND)
%non-destructive
measurements of the qubit’s state via the cavity, leading to so-called dispersive readout of the spin qubit~\cite{DAnjou-Burkard_OptimDispRdOutSpinQB_PRB_2019}.
The main advantage of feedback control is the ability to cope with uncertainty and recover from unexpected events such as noise, measurement errors, and quantum jumps in the case of quantum systems. These are the reasons why feedback protocols have been developed extensively to control classical dynamical systems (with great success). %, and more recently quantum systems. This has motivated the present work.
We hereby present the design of a closed-loop feedback control protocol to drive the qubit towards the desired initial state in a gradual and continuous manner via weak measurements and appropriate control signals applied to the qubit.
Our protocol was motivated by that developed by Haroche’s group to control a quantum cavity using measurements made on Rydberg atoms in the two-level approximation~\cite{dotsenko2009QFB,sayrin2011RTQFBStabPhotNbStates,guerlin2007ProgFieldStateCollapse}, with the significant difference that here the roles are inverted: the two-level system is controlled and the cavity serves for measurements.

In a numerical %comparative
study, we show that our MBQFB protocol achieves higher qubit state initialization fidelities and shorter initialization times than with the current leading approaches mentioned above.
Our approach also has the experimental advantage that it works with the qubit operated at temperatures up to
%2~K
1~K and beyond to a few Kelvins
while maintaining state initialization fidelities over 99.9\%.
%Operation of spin qubits at higher temperature, so-called hot qubits, is currently of high interest as scaling up of spin qubit architectures is under way~\cite{Veldhorst-Dzurak_SiCMOSArchi_2017,Ruoyu-Clarke-Veldhorst_Crossbar_2018,Ruffino-Charbon_CryoCMOS_NatElectron_2022}.
%Scaling up with on chip control electronics that generates heat requires operating the spin qubits at higher temperatures\cite{Vandersypen_ReviewHotDenseCoh_2017,Yang-MPL-Dzurak_OpSiQProcAbove1K_2020,Petit_Intel-QuTech_HotSiQBits_2020,Yoneda-Dzurak_CohSpinTransp_2021,Xue-Vandersypen_Intel-QuTech_CMOSCtrlSi_2021,Camezind_Basel-IBMZurich_HotSpinQubitFinFET4K_2022}.
Current efforts in scaling up spin qubit architectures~\cite{Veldhorst-Dzurak_SiCMOSArchi_2017,Ruoyu-Clarke-Veldhorst_Crossbar_2018,Ruffino-Charbon_CryoCMOS_NatElectron_2022} require on-chip control electronics that generate heat. Hence, initialization of spin qubits at higher temperatures \cite{Vandersypen_ReviewHotDenseCoh_2017,Yang-MPL-Dzurak_OpSiQProcAbove1K_2020,Petit_Intel-QuTech_HotSiQBits_2020,Yoneda-Dzurak_CohSpinTransp_2021,Xue-Vandersypen_Intel-QuTech_CMOSCtrlSi_2021,Camezind_Basel-IBMZurich_HotSpinQubitFinFET4K_2022}, so-called hot qubits, is currently of very high interest.

This paper is organized as follows:
Sect.~\ref{sect:SystModel} presents the mathematical model of the DQD system considered herein comprising a spin and a charge degree of freedom
%, starting from the Hamiltonian and its eigenstates, incorporating its electrical interaction
interacting
with a superconducting cavity.
%, and presenting input-output theory of the compound system considered as an open quantum system.
The spin and charge degrees of freedom of the DQD architecture
%are
can be
reduced to an effective two-level system, allowing to define a qubit and leading to an effective Jaynes-Cummings Hamiltonian describing the interaction between the qubit and the cavity as in cQED.
%Sect.~\ref{sect:Meas}
That section also develops the model for dispersively measuring the qubit via the cavity, enabling weak QND measurements. Such measurements are essential for qubit state initialization relying on MBQFB, which is described in Sect.~\ref{sect:InitQFB}. Results of numerical simulations are presented in Sect.~\ref{sect:Results}, and Sect.~\ref{sect:Conclu} concludes the paper.

\section{System model}
\label{sect:SystModel}

\subsection{Double quantum dot coupled to a cavity}
\label{sect:DQDModel}

 \begin{figure}[t]
		\centering
		\includegraphics[width=\linewidth]{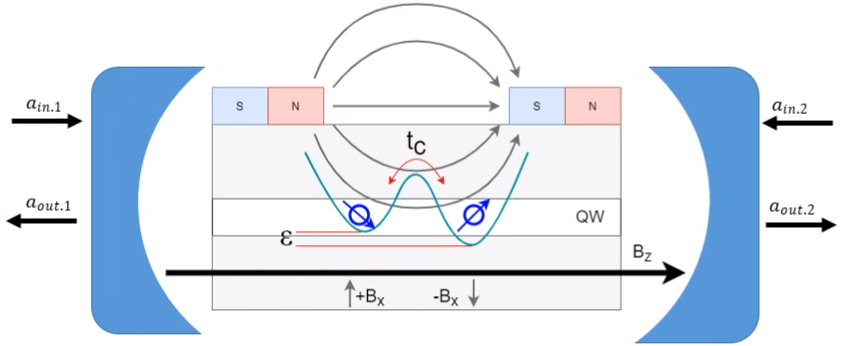}  
		\caption{Schematic representation of a DQD coupled to a cavity, adapted from Benito \textit{et al.}~\cite{benito2017input}.}
		\label{fig:model}
\end{figure}

The system considered herein is that analyzed by Benito \textit{et~al.}~\cite{benito2017input} and realized experimentally in~\cite{mi2018coherent}, comprised of a gate-defined DQD coupled to a microwave superconducting cavity (Fig.~\ref{fig:model}). This DQD-$\mu$WSCc system has a highly desired feature for
%the purpose of
initializing a qubit state by MBQFB as considered here
%as
since
it allows weak
%non-destructive
QND
measurements of the qubit's state via measurement of the cavity's state.

The DQD (Fig.~\ref{fig:model}) confines a single electron in a gate-defined double electrostatic well.
The spatial degree of freedom allows two basis states, for which the electron is located in the left ($L$) or right ($R$) dot (or well), with the possibility of the electron being in a superposition of these basis states. Depending on the applied gate voltages, there is a charging energy difference between the $L$ and $R$ dots to be denoted $\varepsilon$ along with a tunnel coupling energy between the two dots denoted $t_c$. The DQD is subjected to an external longitudinal magnetic field $B_z$ and a small transverse magnetic field gradient $B_x$. The $B_z$ field separates the degenerate energy levels by Zeeman splitting. The  $B_x$ field gradient resulting from a micromagnet fabricated on top of the device induces spin-charge hybridization, which is required to achieve a strong spin-photon coupling with the microwave cavity~\cite{Jin-Schon_etal_PRL108_2012,viennot2015coherent,Mi-Petta_etal_Science_2017}.
Note that in the remainder of the present paper, the magnetic fields $B_z$ and $B_x$ will be given in energy units; hence $B_z$ stands for $g \mu_B B_z$, and similarly for $B_x$, where $g$ is the Land\'{e} factor for the electron ($g = 2$), and $\mu_B$ is the Bohr magneton.
Under appropriate conditions, the DQD-$\mu$WSCc system depicted in Fig.~\ref{fig:model} can be described by the following Jaynes-Cummings Hamiltonian~\cite{benito2017input}:
\begin{equation}
H_\mathrm{JC} = \hbar \omega'_c \, a^\dagger a + \frac{\hbar \omega_q}{2}\sigma_z + \hbar g_s (a^\dagger \sigma_s + a \sigma_s^\dagger) .  
\label{eq:Jaynes-Cummings}
\end{equation}
Appendix~\ref{app:DetailedModel}
%Appendix~A % Pour le mémoire
provides
a summary of the complete model of the DQD-$\mu$WSCc and
the details of the derivation to arrive at the Jaynes-Cummings Hamiltonian along with the expressions for the parameters $\omega_c'$, $\omega_q$, and $g_s$ in terms of the basic parameters of the DQD and the cavity. Here, the operators $a$ and $a^\dagger$ are the annihilation and creation operators for the cavity mode of interest, and the operators $\sigma_z$ and $\sigma_s$ relate to the spin degree of freedom considered as the qubit.

\subsection{Measurements}
\label{sect:Meas}

%In the theory
%%presented above,
%underlying the DQD-$\mu$WSCc (see Appendix~\ref{app:DetailedModel}),
%the Heisenberg picture
%%was
%is
%used. In this section, the Schrödinger picture, in which a state is specified by a time-evolving vector or density operator, will be used instead as it is easier to handle in the context of feedback control.
To
%be able to
implement quantum feedback, it is important that weak,
%non-destructive,
and QND measurements be used to affect as little as possible the state of the qubit as it is being measured. This can be achieved with a qubit-cavity coupling in the dispersive regime, with the effective dispersive Hamiltonian given by~\cite{blais2004cavity}:
\begin{equation}
H_\mathrm{disp} = \hbar (\omega'_c + \chi \sigma_z) \, a^\dagger a + \frac{\hbar (\omega_q + \chi)}{2}\sigma_z ,
\end{equation}
with $\chi = g_s^2 / (\omega'_c - \omega_q)$ being the dispersive interaction strength. %\mid \tilde{\omega}_c - \omega_q \mid$.
This Hamiltonian is an approximation of the Jaynes-Cummings model in the case of a large frequency detuning between the qubit and the cavity relative to the coupling strength~$g_s$~\cite{haroche2006exploring}.

\begin{figure}[ht]
	\centering
	\includegraphics[width=\linewidth]{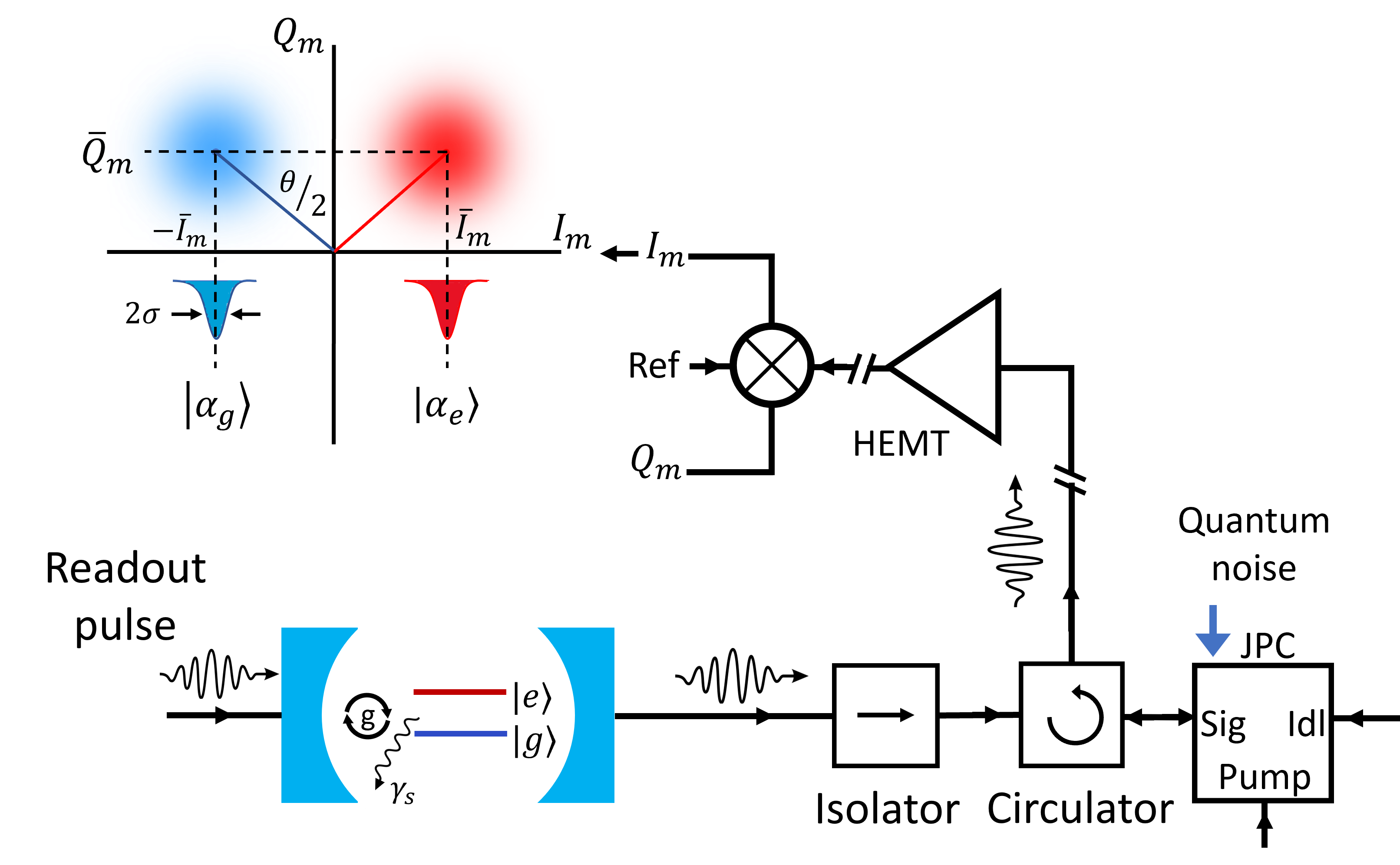}  
	\caption{Schematic illustration of the measurement chain of the DQD via a cavity. Adapted from~\cite{hatridge2013VarStrengthMeas}.}
	\label{fig:mesure}
\end{figure}

The two physical qubit states $\ket{g}$ and $\ket{e}$ serve as the logical qubit states. To measure a qubit in a general state $\ket{\psi} = c_g\ket{g} + c_e\ket{e}$, the cavity is first emptied and initialized in a coherent state $\ket{\alpha}$ (the $\alpha$ here shall not be confused with that of Eq.~\eqref{eq:alphaQubitDQD}). The qubit and cavity are then entangled for a duration $T_m$ under the unitary evolution of the dispersive Hamiltonian; $T_m$ defines the measurement time. The combined qubit-cavity system state after the entanglement interaction is:
\begin{equation}
\ket{\Psi} = c_g\ket{g} \otimes \ket{\alpha_g}
           + c_e\ket{e} \otimes \ket{\alpha_e} .
\label{eq:StateQubitCavAfterEntangl}
\end{equation}
This shows that it is possible to measure the state of the cavity after entanglement with the qubit and subsequently infer the qubit's state; the latter being thus measured indirectly via the cavity. Experimentally, a Josephson parametric converter (JPC) device can be used to perform such a measurement and the remainder of this section is an adaptation for the present purposes of some of the developments presented in Ref.~\cite{hatridge2013VarStrengthMeas}.
Fig.~\ref{fig:mesure} depicts a measurement chain using a JPC.
The JPC allows recording two output values $(I_m, Q_m)$, which are used to determine the new qubit state after a measurement. The $\alpha$'s of the coherent states appearing in Eq.~\eqref{eq:StateQubitCavAfterEntangl} are related to the means of the output values $\overline{I}_m$ and $\overline{Q}_m$ as follows~\cite{hatridge2013VarStrengthMeas}:
\begin{gather}
  \alpha_g = -\overline{I}_m + i \overline{Q}_m , \\
  \alpha_e = \overline{I}_m + i \overline{Q}_m,
\end{gather}
whereby $\overline{I}_m$ and $\overline{Q}_m$ are related to each other
and can be determined
through the following relationships~\cite{hatridge2013VarStrengthMeas}:
\begin{equation}
\overline{I}_m^{\,\, 2} +\overline{Q}_m^{\, 2} = \bar{n} \kappa T_m ,
\quad
\frac{\overline{I}_m}{\overline{Q}_m}=\frac{\chi}{\kappa},
\end{equation}
with $\kappa$ being the cavity decay rate, and $\bar{n} = \frac{|\alpha|^2}{\kappa T_m}$. % ($|\alpha|^2 = |\alpha_g|^2 = |\alpha_e|^2$).
Using the density operator to specify the state of the qubit,
%obtained by tracing out the cavity,
and denoting that density operator before a measurement by $\rho^q_\mathrm{before}$, the state after measurement of the cavity delivering $(I_m, Q_m)$ is~\cite{hatridge2013VarStrengthMeas}
\begin{equation}
  \rho^q_\mathrm{after}(I_m,Q_m)
  =
  \frac{M_{I_m,Q_m} \, \rho^q_\mathrm{before} \, M_{I_m,Q_m}^{\dagger}}
       {\Tr(M_{I_m,Q_m} \, \rho^q_\mathrm{before} \, M_{I_m,Q_m}^{\dagger})} ,
\end{equation}
with
\begin{equation}
M_{I_m,Q_m}
=
\frac{e^-\frac{(Q_m-\overline{Q}_m)^2}{4\sigma^2_m}}{\sqrt{\pi}}
\begin{bmatrix}
  e^{-\frac{(I_m - \overline{I}_m)^2}{4\sigma^2_m}} e^{i \frac{\overline{I}_m Q_m}{2\sigma^2_m}}
  & 0  \\
  0 &
  e^{-\frac{(I_m + \overline{I}_m)^2}{4\sigma^2_m}} e^{-i \frac{\overline{I}_m Q_m}{2\sigma^2_m}}
\end{bmatrix} ,
\label{eq:MeasOp}
\end{equation}
where $\sigma^2_m = \frac{1}{2}$ is the fundamental quantum noise associated with a measurement in both quadratures $I$ and $Q$.

Classical noise is also present in measurements and must be considered in addition to the fundamental quantum noise. For the measurements considered here, classical noise can be assumed to be Gaussian and zero mean with variance $\sigma_c^2$ in both quadratures~\cite{hatridge2013VarStrengthMeas}. Introducing the total observed variance $\sigma^2 = \sigma_c^2 + \sigma_m^2$ after an imperfect (so-called inefficient) measurement due to the classical noise, and defining the measurement efficiency as $\eta_m = \sigma_m^2 / \sigma^2$, the qubit density operator after an inefficient measurement that delivers measurement values $I_m$ and $Q_m$ becomes
\begin{equation}
\rho^q_\mathrm{after}(I_m,Q_m)
=
  \frac{\int\int dI dQ \, P(I_m,Q_m|I,Q) \, M_{I,Q} \, \rho^q_\mathrm{before} \, M_{I,Q}^\dagger}
       {\Tr\left(\int\int dI dQ \, P(I_m,Q_m|I,Q) \, M_{I,Q} \, \rho^q_\mathrm{before} \, M_{I,Q}^\dagger\right)} .
\label{eq:EvolQubitStateAfterMeasImQm}
\end{equation}
Here the measurement operator $M_{I,Q}$ is that of Eq.~\eqref{eq:MeasOp} with $I_m$ and $Q_m$ replaced by $I$ and $Q$ (note, however, that the means $\overline{I}_m$ and $\overline{Q}_m$ remain as such in the expression of $M_{I,Q}$, as well as $\eta_m$),
%  \textcolor{red}{******** DIRE CE QU'EST LA MATRICE $M_{I,Q}$ CAR CE N'EST PAS CLAIR, EST-CE $M_{I_m,Q_m}$ AVEC $I_m$ ET $Q_m$ REMPLACÉES PAR $I$ ET $Q$ TOUT EN LAISSANT $\overline{I}_m$ ET $\overline{Q}_m$; DOIT ON REMPLACER $\sigma_m$ PAR AUTRE CHOSE, OU BIEN LAISSER $\sigma_m$ TEL QUEL? JE PENSE QU'IL FAUT LAISSER $\sigma_m$ TEL QUEL, CAR IL S'AGIT DE L'OPÉRATEUR DE MESURE QUANTIQUE. LE BRUIT CLASSIQUE EST INCLUS DANS $P(I_m,Q_m|I,Q)$  ***************}
and~\cite{hatridge2013VarStrengthMeas}
\begin{equation}
P(I_m,Q_m|I,Q)=\frac{\exp-\frac{(I_m-I)^2}{2(1-\eta_m)\sigma^2}\exp-\frac{(Q_m-Q)^2}{2(1-\eta_m)\sigma^2}}{2\pi(1-\eta_m)\sigma^2} .
\end{equation}
The latter probability density accounts for the classical measurement noise ($\sigma_c^2 = \sigma^2 - \sigma_m^2 = (1-\eta_m)\sigma^2$). A value $\eta_m = 0.6$ will be used in what follows, which is experimentally realistic~\cite{hatridge2013VarStrengthMeas}.

\begin{figure}[ht]
	%\hfill
	\subfigure[$T_m=200$~ns.]{\includegraphics[width=4.2cm]{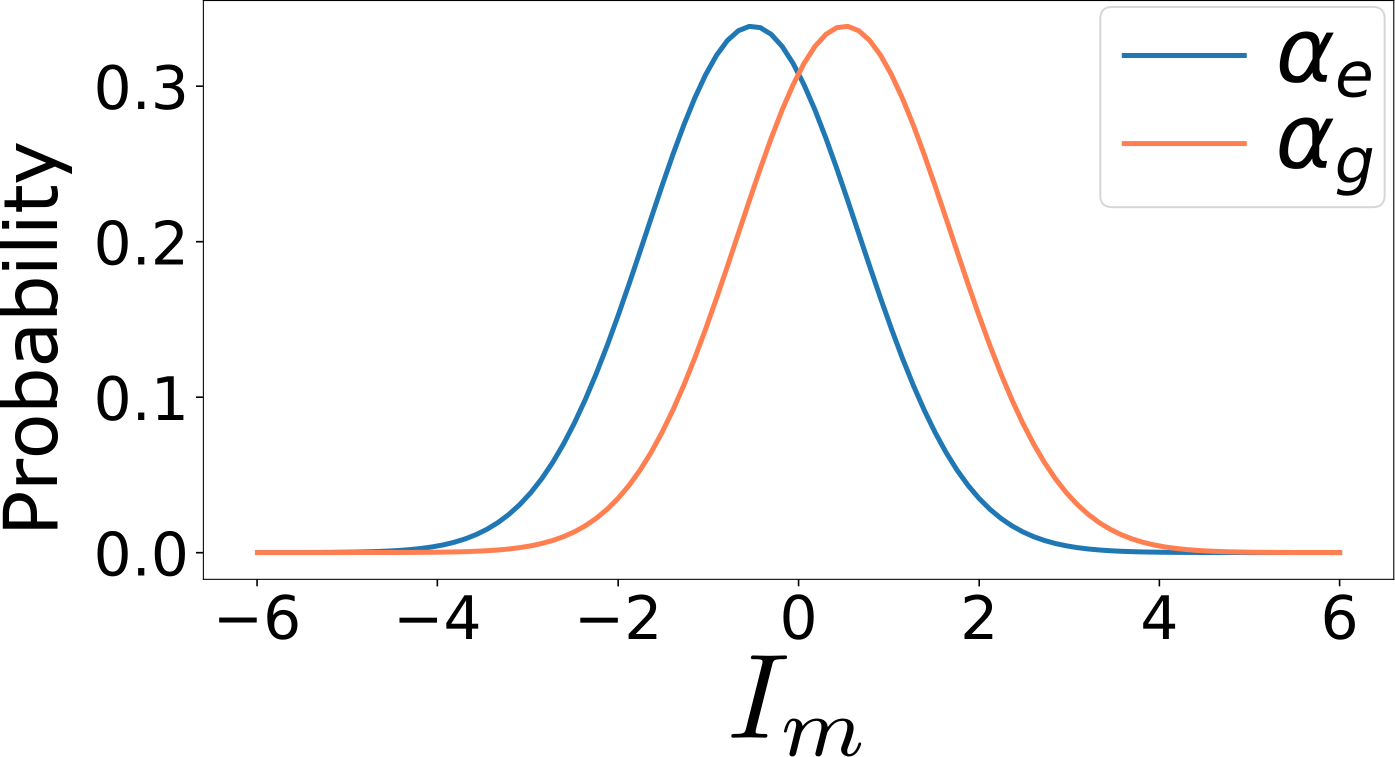}}
	%\hfill
	\subfigure[$T_m=2~\mu$s.]{\includegraphics[width=4.2cm]{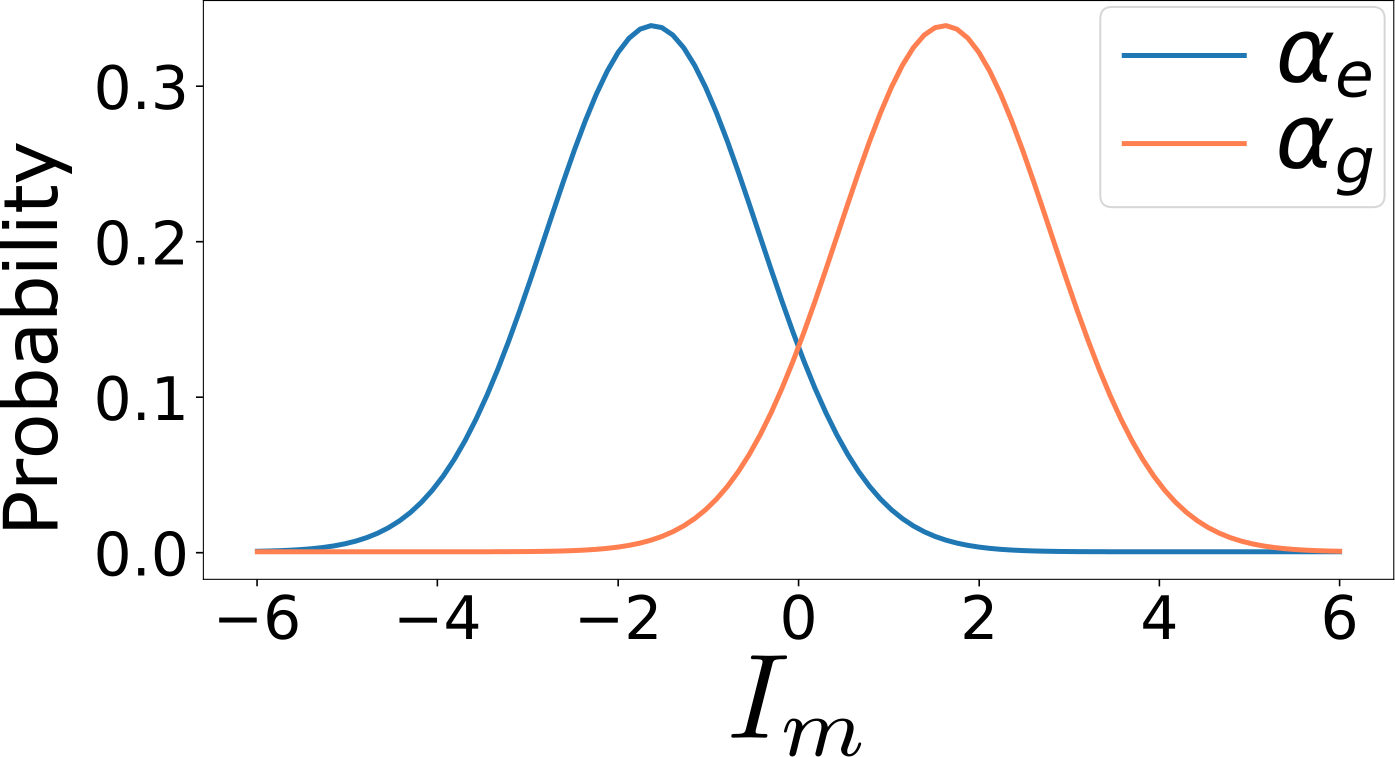}}
	%\hfill
	\caption{Gaussian distributions  $N_{\alpha_e}$ and $N_{\alpha_g}$.}
	\label{fig:deuxdistributions}
\end{figure}

Including classical noise, the measurement values $I_m$ and $Q_m$ are Gaussian distributed with means $\overline{I}_m$ and $\overline{Q}_m$ and variance $\sigma^2$. When the two Gaussian distributions $N_{\alpha_e}$ and $N_{\alpha_g}$ overlap, a weak measurement can be performed on the qubit, whereas two distributions that are well separated leads to a projective measurement of the qubit. The entangling time $T_m$ is the important parameter controlling the overlap between the two Gaussian distributions, and therefore the measurement strength, see
Fig.~\ref{fig:deuxdistributions} for two illustrative cases. For $T_m=200$~ns, the overlap between the two distributions is large. In this case, when the qubit measurement is indirectly performed via the cavity, the qubit state is less disturbed and only a small amount of information about this state is extracted. For $T_m = 2$~$\mu$s, the overlap between the two distributions is small (distributions apart), and when a measurement is performed, the qubit state is thereby strongly disturbed. In this case, the qubit after the measurement is in either the $\ket{g}$ or the $\ket{e}$ state and the quantum information encoded within the qubit state prior to the measurement is destroyed;
% for subsequent use;
this is evidently not desirable for feedback control.

\begin{figure}[ht]
	%\hfill
	\subfigure[Case without decoherence.]{\includegraphics[width=4.2cm]{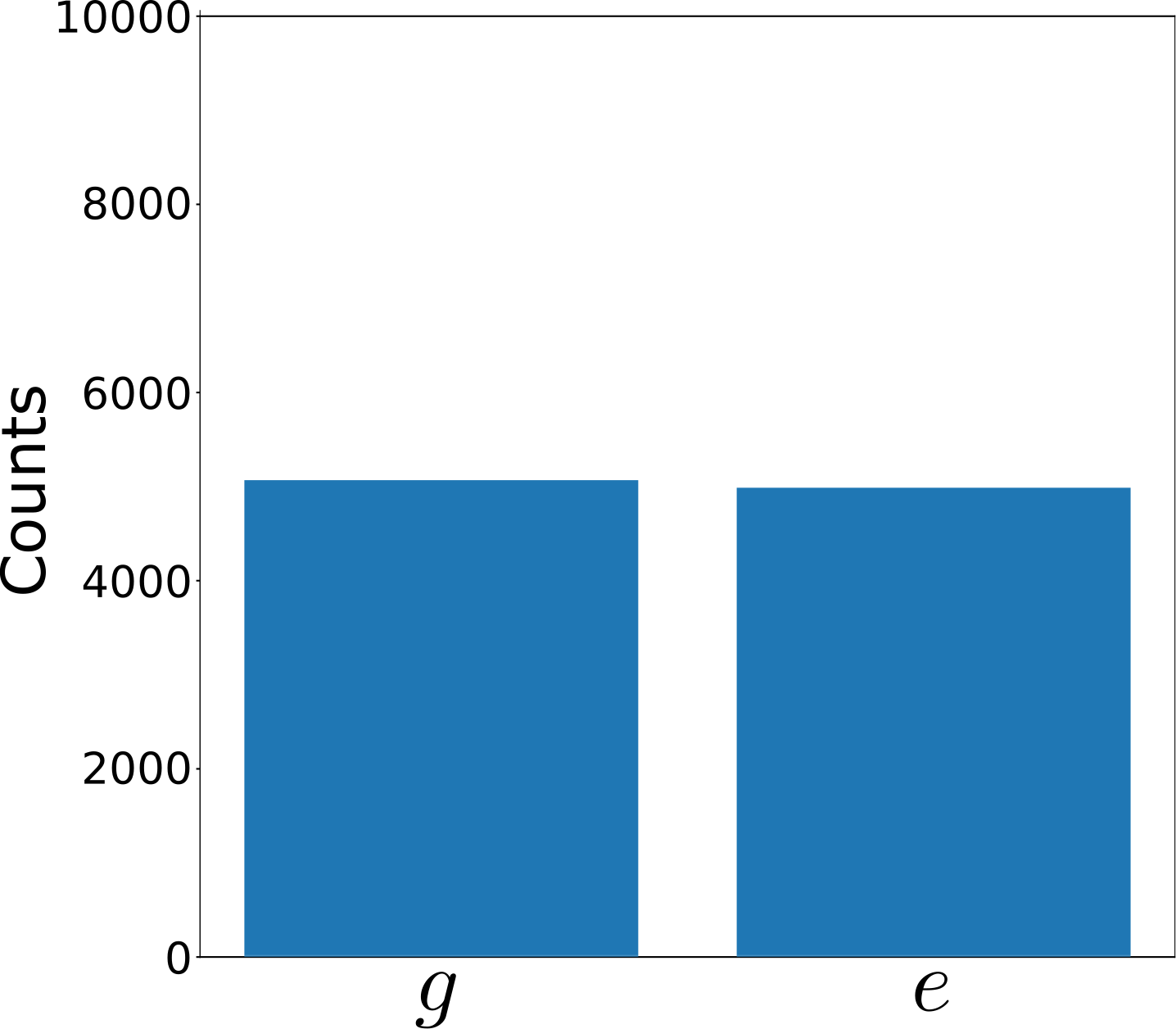}}
	%\hfill
	\subfigure[Case with decoherence. ]{\includegraphics[width=4.2cm]{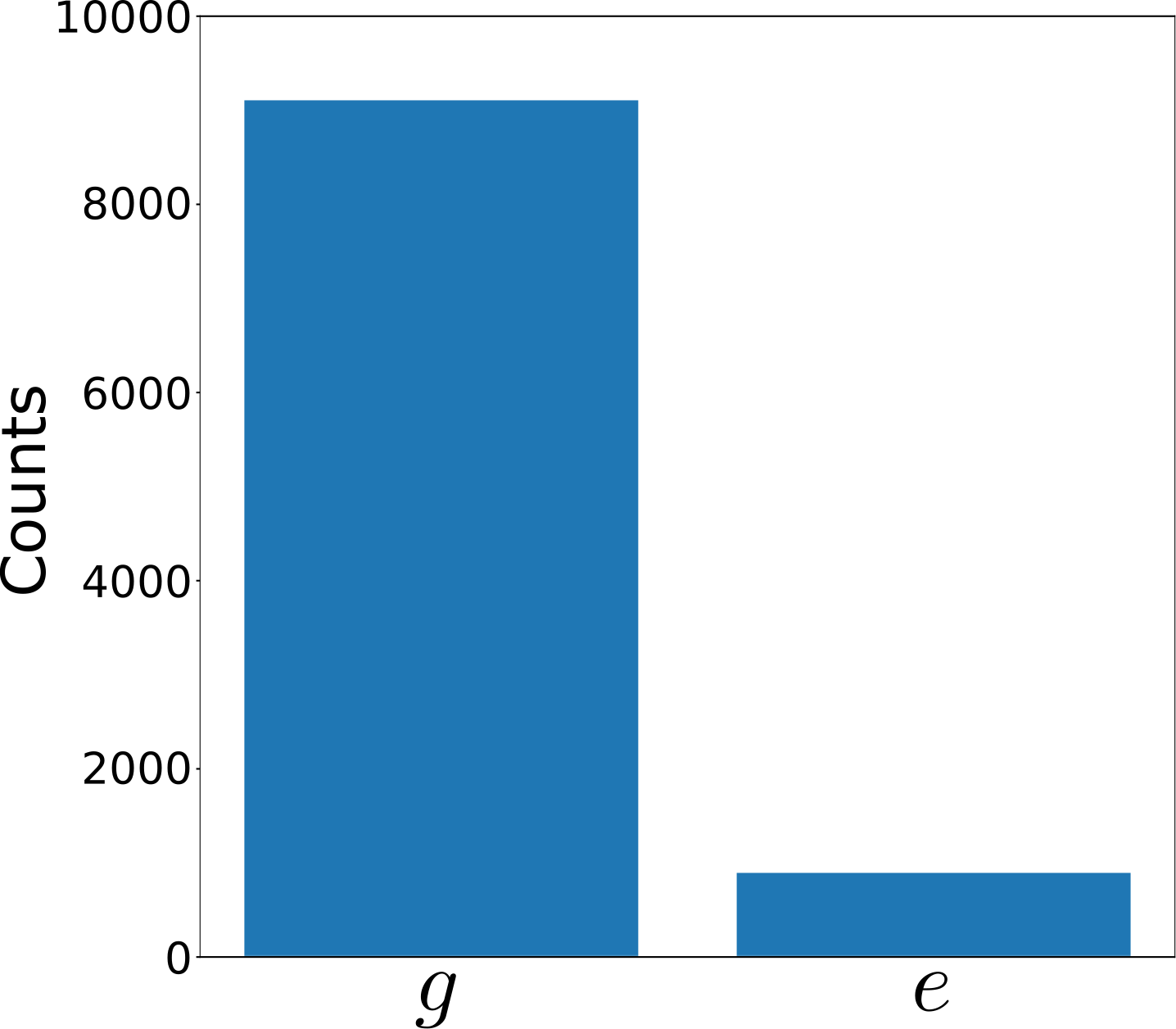}}
	%\hfill
	\caption{Distribution of the states $\ket{g}$ and $\ket{e}$ in the validation process of the measurement operator.}
	\label{fig:validationoperateurmesure}
\end{figure}

The diagonal form of the measurement operator $M_{I_m, Q_m}$ with unequal diagonal elements shows that only the states $\ket{g}$ and $\ket{e}$ can be prepared by way of quantum feedback as it converges. The reason is that upon convergence, the qubit is repeatedly measured with very little control applied to it. Hence, only eigenstates of the measurement operators can be prepared with quantum feedback~\cite{dotsenko2009QFB,sayrin2011RTQFBStabPhotNbStates,guerlin2007ProgFieldStateCollapse}. Fig.~\ref{fig:validationoperateurmesure} shows results validating this in the present case. The validation process consists of measuring the qubit state in sequence a number of times defined as $n_{rep}$. When $n_{rep}$ is large ($n_{rep} = 100$ is used here), the qubit ends up in either the state $\ket{g}$ or $\ket{e}$.
The histograms in Fig.~\ref{fig:validationoperateurmesure} were obtained with 10,000 runs of measurements,
taking the final state of the qubit at the end of
each run containing $n_{rep}$ measurements.
Fig.~\ref{fig:validationoperateurmesure}~(a) illustrates the case with no decoherence. The states $\ket{g}$ and $\ket{e}$ are equiprobable after measurement in this situation. Fig.~\ref{fig:validationoperateurmesure}~(b) depicts the results in the presence of decoherence (cavity and qubit). In this case, owing to decoherence, the qubit state is much more likely to be found in $\ket{g}$ after measurement.
%It can therefore be concluded that the $M$ operator is valid. 
 

\section{Qubit state initialization using quantum feedback control}
\label{sect:InitQFB}

\begin{figure*}[t!]
	\centering
	\includegraphics[width=0.7\linewidth]{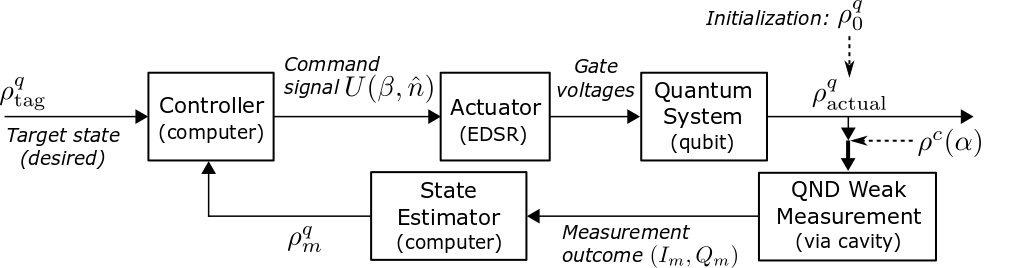} % 0.7
	\caption{Block diagram of the feedback protocol.}
	\label{fig:MBQFBQubitBlockDiag}
\end{figure*}

The goal of control is to find a command signal to bring a system (the qubit here) to a desired state.
In feedback control, this is achieved by monitoring the state of the system during control and adapting the command signal accordingly %as necessary
%all along
to reach the desired state.

As discussed above in Sect.~\ref{sect:DQDModel}, the equivalent two-level system serving as the qubit interacting with the cavity can be described by a Jaynes-Cummings Hamiltonian. A strong analogy can thus be established with cavity quantum eletrodynamics (CQED) and its analogon cQED with circuits. Hence control techniques inspired from those developed for CQED and cQED can be adapted.
To control the qubit, it is assumed here that a general unitary gate having the form
\begin{equation}
  U(\beta, \hat{n}) = \exp\left( -i \, \frac{\beta}{2} \, \hat{n} \scaldot \vec{\sigma} \right)
                    = \cos\frac{\beta}{2} I - i\sin\frac{\beta}{2} \hat{n}\scaldot\vec{\sigma}
\end{equation}
can be applied to its state, where $\beta$ is the "spin rotation" angle and $\hat{n} = n_x \vec{\imath} + n_y \vec{\jmath} + n_z \vec{k} \equiv (n_x, n_y, n_z)$ is a unit vector ($\nrm{ \hat{n}} = 1$) specifying the axis around which the spin rotation takes place ($\vec{\imath}$, $\vec{\jmath}$ and $\vec{k}$ are the unit vectors along the coordinates axes $x$, $y$ and $z$). Hence, $\beta$ and the direction specified by $\hat{n}$
%, which can be parametrized using spherical angles as $\hat{n} = (\sin\theta\cos\phi, \sin\theta\sin\phi, \cos\theta)$
are the available parameters for control. In effect, the vector $\beta \hat{n}$ contains 3 independent parameters, and as will be seen below, it is those parameters that will be chosen appropriately for control.
Such a unitary %gate
can be realized by way of electric dipole spin resonance (EDSR). For the DQD architecture considered here, EDSR can be implemented by subjecting the electron in the DQD to a longitudinal magnetic field ($B_z$ in Fig.~\ref{fig:model}) and using gate voltages in the radio-frequency range applied to high frequency electrodes to move the electron back and forth from one dot to the other in the DQD, which makes the electron see an alternating transverse radio-frequency magnetic field along $x$ as needed for EDSR~\cite{mpl2007MicroMagCohCtrl,Croot-Burkard-Petta_FloppingModeEDSR_PRRes_2020}.

Let $\rho^q_m$ be the state of the qubit after a measurement and prior to applying the control via the gate $U$. Then, the state of the qubit $\rho^q_c$ after applying the control is given by
\begin{equation}
  \rho^q_c = U(\beta, \hat{n}) \rho^q_m U^\dagger(\beta, \hat{n}) .
\label{eq:StateQubitAfterCtrl}
\end{equation}
One may object that if a general unitary gate is possible, then reaching any desired state with such a gate will be possible, and hence there is no need for feedback control. This is
% too simplistic a
a simplistic
view because this would require that the qubit be prepared in an \textit{a priori} known state before applying the gate, and that is exactly the problem addressed here, namely preparing that initial state. Furthermore, if an unexpected event occurs during state preparation, then there can be no guarantee that the desired state will be reached. Feedback control as considered here will thus consist in gradually bringing the state in a controlled way to the desired end state by monitoring its state during state preparation, %along the way
and adapting the command signal as necessary. This is similar in spirit to the protocol devised by Serge Haroche’s group to prepare Fock states in a microwave cavity~\cite{dotsenko2009QFB,sayrin2011RTQFBStabPhotNbStates} as part of a CQED set-up. As mentioned previously, in the situation considered here, the cavity is used to
%obtain
gain
information %knowledge
%increase the state of knowledge
about the qubit's state in a
%non-destructive
QND
manner and by least disturbing
%it.
that state.
%Then, it is possible to follow the evolution of the qubit state and prepare it toward the target state.

The control objective is to ultimately reach a state that has the smallest fidelity distance from the target % desired
state to be denoted $\rho^q_\mathrm{tag}$. The fidelity distance $d_F(\rho_1, \rho_2)$ between two states $\rho_1$ and $\rho_2$ used here is defined as $d_F(\rho_1, \rho_2) = 1 - F(\rho_1, \rho_2)$, where $F(\rho_1, \rho_2) = \Tr(\rho_1 \rho_2)$ is the fidelity.
The control problem can thus be cast as minimizing the fidelity distance. Such minimization will be carried out iteratively by requiring that for each feedback loop (iteration), and given the information obtained about the state via a measurement made in the loop, the distance between the state $\rho^q_c$, after the $U$ gate is applied, and the target state $\rho^q_\mathrm{tag}$ be smaller than the distance between the state prior to the application of the $U$ gate, \textit{i.e.} the state after measurement $\rho^q_m$, and the target state, that is
\begin{equation}
  d_F(\rho^q_c, \rho^q_\mathrm{tag}) < d_F(\rho^q_m, \rho^q_\mathrm{tag}),
\end{equation}
which, using Eq.~\eqref{eq:StateQubitAfterCtrl},
%is equivalent
amounts
to
\begin{equation}
  \Tr(U \rho^q_m U^\dagger \rho^q_\mathrm{tag}) > \Tr(\rho^q_m \rho^q_\mathrm{tag}) .
\label{eq:CondFB}
\end{equation}
By developing to first order the left-hand side of the previous inequality for small values of $\beta$,
% see Appendix~\refA JE PENSE QU'IL FAUDRAIT DONNER CE CALCUL DANS UNE ANNEXE ET LA CITER ICI
in which case $U(\beta, \hat{n}) \approx I - i \, \frac{\beta}{2} \hat{n}\scaldot\vec{\sigma}$,
it is shown that this inequality is satisfied by chosing $\beta \hat{n}$, such that
\begin{equation}
\beta \big( n_x \Tr\big( i \sigma_x C \big)
          + n_y \Tr\big( i \sigma_y C \big)
          + n_z \Tr\big( i \sigma_z C \big)
      \big) > 0 ,
\label{eq:CondFBAngleAndAxisRot}
\end{equation}
where $C$ stands for the commutator given by $C = [\rho_\mathrm{tag}^q, \rho_m^q]$.
Defining the vector
\begin{equation}
\vec{v} = \Tr\big( i \sigma_x C \big) \vec{\imath}
        + \Tr\big( i \sigma_y C \big) \vec{\jmath}
        + \Tr\big( i \sigma_z C \big) \vec{k} ,
\end{equation}
the condition in Eq.~\eqref{eq:CondFBAngleAndAxisRot} is equivalent to
$\beta \hat{n} \scaldot \vec{v} > 0$.
For this condition to be satisfied, $\beta$ can be chosen positive and small, and $\hat{n}$ can be taken as $\hat{n} = \vec{v} / \nrm{\vec{v}}$.
More specifically as regards $\beta$, its value must be chosen sufficiently small such that when developing the left-hand side of the inequality in Eq.~\eqref{eq:CondFB} in powers of $\beta$, the second order term is much smaller than the first order term to ensure that the inequality holds robustly to first order (for this calculation, $U(\beta, \hat{n})$ must be expanded to second order in~$\beta$). This leads to the following condition on $\beta$:
\begin{equation}
  \beta
  \ll
  \zeta,
  \, \mathrm{with} \,\,\,
  \zeta
  =
  \left|
  \frac{\Tr\big(\vec{n} \scaldot \vec{\sigma} \, C \big)}
       {\Tr\big( \vec{n} \scaldot \vec{\sigma}
                 \rho_m^q
                 \vec{n} \scaldot \vec{\sigma}
                 \rho_\mathrm{tag}^q \big)
        - \Tr\big( \rho_m^q \rho_\mathrm{tag}^q \big)}
  \right| .
\end{equation}
In practice, for the results presented below %, the value of $\beta$ is set to
$\beta = \epsilon \zeta$, with $\epsilon = 0.1$ (the exact value of $\epsilon$ does not affect much the results, $\epsilon = 0.01$ has also been tested without significant changes).
%$\beta = 0.1 \times \zeta$.

The feedback protocol of the qubit considered here is implemented in an iterative process similar to that presented in~\cite{dotsenko2009QFB}, and is depicted as a block diagram in Fig.~\ref{fig:MBQFBQubitBlockDiag}.
In this figure, the actuator is the physical interface that translates a computed command signal %at its input
into a physical action that can physically be applied (typically in the form of fields) to the quantum system (qubit here) to control it. More specifically here, this physical interface consists of the radio-frequency voltages applied to the high frequency electrodes that allow EDSR.
The
%process
feedback protocol
starts from an initial \textit{a priori} unknown qubit state $\rho^q_0$. As indicated in Fig.~\ref{fig:MBQFBQubitBlockDiag}, this corresponds to $\rho^q_\mathrm{actual}$ being initialized to $\rho^q_0$.
In each feedback loop (iteration), prior to measurement, the cavity is emptied and initialized to a coherent state $\rho^c(\alpha) = \ket{\alpha}\bra{\alpha}$ as described in Section~\ref{sect:Meas} (see also Fig.~\ref{fig:MBQFBQubitBlockDiag}). The combined qubit-cavity state at this point is
$\rho^{q,c}_\mathrm{actual} = \rho^q_\mathrm{actual} \otimes \rho^c(\alpha)$. To account for decoherence and thermal events affecting both the qubit and cavity, the free evolution of the combined state is modeled
%according to
using
a Lindblad master equation. In a real experiment this free evolution occurs by itself, but for the
purpose %sake
of feedback control, it is also needed to model this evolution in order to obtain an estimate of the state for calculating the command signal.
%More precisely, the
The Lindblad equation accounts for: (i) the decoherence of the qubit (longitudinal and transverse relaxation), (ii) thermal excitation and relaxation of the qubit through interaction with the thermal environment at temperature $T$, (iii) relaxation of the cavity, and (iv) thermal excitation and relaxation of the cavity with the thermal environment.
In the numerical modelling, this evolution
%is considered to take
takes
place prior to measurement. This way of proceeding assumes that this free evolution accounts for all decoherence and thermal processes that may occur during a feedback loop iteration.
Free evolution is defined here as evolution not caused by measurement back-action (Eq.~\eqref{eq:EvolQubitStateAfterMeasImQm}) or the action of control (Eq.~\eqref{eq:StateQubitAfterCtrl}). A further underlying assumption for the sake of modelling is that no decoherence or thermal events occur while performing a measurement of the cavity, or during control action of the qubit. These assumptions, which are customary, ease the modelling process,
% by segmenting it into individual processes,
although in reality, there can be decoherence and thermal events during a cavity measurement, or a control action. Such assumptions nevertheless
provide for accurate and realistic modelling results as shown by the work of Haroche \textit{et al.} in the case of controlling a microwave quantum cavity~\cite{dotsenko2009QFB,sayrin2011RTQFBStabPhotNbStates}, and
%, which
are justified by the fact that decoherence and thermal events are stochastic. Hence, realistic modelling results are obtained by assuming that these events occur during a finite period of time in the course of a feedback loop, and apart from cavity measurement and qubit control action.
The Lindblad equation for the combined (entangled) evolution of the qubit and cavity prior to measurement of the cavity in each feedback loop is given by:
\begin{equation}
\begin{split}
  \frac{\ddif \rho^{q,c}}{\ddif t}
  %\dot{\rho}^{q,c}
    &=
  -i [ H_\mathrm{JC}, \rho^{q,c} ]
  \\
  &\relphantom{=} {} - \frac{\kappa'(1 +  n_\mathrm{th}(\omega_c))}{2}
    ( a^\dagger a  \rho^{q,c} +  \rho^{q,c} a^\dagger a - 2 a  \rho^{q,c} a^\dagger )
  \\
  &\relphantom{=} {} - \frac{\kappa' n_\mathrm{th}(\omega_c)}{2}
    ( a a^\dagger \rho^{q,c} + \rho^{q,c} a a^\dagger - 2 a^\dagger \rho^{q,c} a )
  \\
  &\relphantom{=} {} - \frac{\gamma_s (1 +  n_\mathrm{th}(\omega_q))}{2}
    ( \sigma_s^\dagger \sigma_s  \rho^{q,c}
      + \rho^{q,c} \sigma_s^\dagger \sigma_s - 2 \sigma_s  \rho^{q,c} \sigma_s^\dagger )
  \\
  &\relphantom{=} {} - \frac{\gamma_s n_\mathrm{th}(\omega_q)}{2}
    ( \sigma_s \sigma_s^\dagger \rho^{q,c}
      + \rho^{q,c} \sigma_s \sigma_s^\dagger - 2 \sigma_s^\dagger \rho^{q,c} \sigma_s ) ,
\end{split}
\label{eq:LindbladEvolQubitCav}
\end{equation}
with $\rho^{q,c}(0) = \rho^q_\mathrm{actual} \otimes \rho^c(\alpha)$, and where
\begin{equation}
  n_\mathrm{th}(\omega) = \frac{1}{e^{\hbar\omega / k_B T} - 1}
\end{equation}
is the average number of thermal photons per mode at frequency $\omega$ given by Planck's law. % ($k_B$ is Boltzmann's constant).
The parameters used in $H_\mathrm{JC}$ appearing in Eq.~\eqref{eq:LindbladEvolQubitCav} are those of the dispersive regime.
%and whereby this evolution occurs for a time duration $T_m$.
The qubit and cavity evolve according to the previous Lindblad equation for a duration $T_m = 200$~ns (duration of a weak measurement as seen above).
Following this evolution, the qubit is subjected to a QND weak measurement via an ($I, Q)$ measurement of the cavity as previously described in Sect.~\ref{sect:Meas}.
%In the simulation, the cavity is traced out and   IL N'Y A PAS DE TRACE PARTIELLE, CAR LE FAIT DE MESURER LA CAVITÉ FAIT UNE PROJECTION SUR L'ESPACE DES ÉTATS DU QUBIT
The state of the qubit $\rho^q_m$ indicated in Fig.~\ref{fig:MBQFBQubitBlockDiag} and following a measurement outcome $I_m$ and $Q_m$ is modeled using Eq.~\eqref{eq:EvolQubitStateAfterMeasImQm}.
This state is then compared by the controller against the target state using the fidelity distance. If this distance is small enough (\textit{i.e.} smaller than a preset threshold), then the target state is considered to be reached. Otherwise, the controller computes the parameters $\beta$ and $\hat{n}$ of the control operator $U(\beta, \hat{n})$ as described above in order to reduce the fidelity distance. The operator $U(\beta, \hat{n})$ is then applied to the qubit by way of EDSR and another feedback iteration is initiated.

In this paper, the control protocol is developed and tested by way of numerical simulations. In a real experiment, the sequence of modelling steps within a given feedback iteration described above defines a quantum filter~\cite{Bouten-VanHandel_IntroQFilt_SIAMJCtrlOptim_2007,Bouten-VanHandel_DiscreteInvitQFiltFBCtrl_SIAMReview_2009}. During an experiment, the exact state of the qubit cannot be known. However, a quantum filter, which provides a state estimator analogous to a Kalman filter is classical control theory, allows having a real-time estimate of the state of the qubit in the computer. This estimate enables computing the parameters $\beta$ and $\hat{n}$ of the control operator $U(\beta, \hat{n})$ to be applied at the end of the iteration as demonstrated by Haroche \textit{et al.}~\cite{sayrin2011RTQFBStabPhotNbStates}.
Athough prior to feedback the state of the qubit is unknown,
accumulation of information on the qubit state via weak measurement outcomes over multiple feedback loops allows the filter to provide a faithful and reliable real-time representation of the qubit's real state irrespective of the initial state~\cite{dotsenko2009QFB}.
%Indeed, it can be proved that as the number of iterations goes to infinity, the estimated state converges to a state independent of the initial state~\textcolor{red}{À VALIDER; CITER ICI UN ARTICLE SUR FILTRAGE QUANTIQUE}.

For clarity, each of the elementary operations on the qubit state (Lindblad evolution, weak measurement, and control) will be represented in what follows using the superoperator formalism~\cite{Wiseman_Milburn_BookQMeasCtrl_2009,dotsenko2009QFB}. Hence, assuming $\rho^q_\mathrm{before}$ is the qubit state prior to an elementary operation and $\rho^q_\mathrm{after}$ is the state after that operation, then the general form of the equation representing the effect of that operation on state $\rho^q_\mathrm{before}$ to give state $\rho^q_\mathrm{after}$ will generally be written as $\rho^q_\mathrm{after} = \mathbf{\mathsf{S}} \rho^q_\mathrm{before}$, where $\mathbf{\mathsf{S}}$ is the superoperator representing the operation. Using this formalism, the evolution of the qubit state resulting from the Lindbladian evolution during a time $T_m$ (Eq.~\eqref{eq:LindbladEvolQubitCav}) is written as $\rho^q_\mathrm{after} = \mathbf{\mathsf{L}}(T_m) \rho^q_\mathrm{before}$.
%whereby $\mathsf{L} \rho$ represents the evolution of a state $\rho$ under the Lindblad equation.
Likewise, Eq.~\eqref{eq:EvolQubitStateAfterMeasImQm} for a measurement with outcome $I_{m}$, $Q_{m}$, will be written as $\rho^q_\mathrm{after} = \mathbf{\mathsf{M}}_{I_m, Q_m} \rho^q_\mathrm{before}$. Finally, the action of the control unitary gate (Eq.~\eqref{eq:StateQubitAfterCtrl}) takes the form
$\rho^q_\mathrm{after} = \mathbf{\mathsf{U}}(\beta, \hat{n}) \rho^q_\mathrm{before}$.
Summarizing, and provided that the state of the qubit is $\rho_{\mathrm{begin}, j}^q$ at the beginning of the $j$-th iteration, the state of the qubit at the end of the iteration is given by:
\begin{equation}
  \rho_{\mathrm{end}, j}^q
  =
  \mathbf{\mathsf{U}}(\beta_j, \hat{n}_j) \,
  \mathbf{\mathsf{M}}_{{I_{m_j}}, Q_{I_{m_j}}}
  \mathbf{\mathsf{L}}_j(T_m) \,
  \rho_{\mathrm{begin}, j}^q \, .
\end{equation}
This equation can be applied recursively at each iteration to get a real-time estimate of the state of the qubit.
Starting from an initial qubit state $\rho^q_0$ prior to feedback, the estimate of the state provided by the quantum filter at the end of iteration~$k$ ($k = 0$ corresponding to the initial state) is given by
\begin{equation}
  \rho_{\mathrm{end}, k}^q
  =
  \prod_{j=1}^k
  \mathbf{\mathsf{U}}(\beta_j, \hat{n}_j) \,
  \mathbf{\mathsf{M}}_{{I_{m_j}}, Q_{I_{m_j}}}
  \mathbf{\mathsf{L}}_j(T_m) \,
  \rho_0^q \, .
\end{equation}
As mentioned, feedback control loops are iteratively applied until convergence, \textit{i.e.} until the fidelity distance reaches a value below a predetermined threshold value.

\section{Results}
\label{sect:Results}

The target state for initialization is the qubit's ground state $\ket{g}$ considered to be the logical state $\ket{0^q}$. Note that given the qubit is in a state $\rho^q$, the probability of preparing the qubit's ground state $\ket{\psi^q_\mathrm{tag}} = \ket{0^q}$, with corresponding density operator $\rho^q_\mathrm{tag} = \ket{0^q} \bra{0^q}$, is the same as the fidelity since $F(\rho^q, \rho^q_\mathrm{tag}) = \Tr(\rho^q \rho^q_\mathrm{tag}) = \Tr\big(\rho^q \ket{0^q} \bra{0^q}\big) = \wp(0)$.

\vspace{-20pt}

\begin{table}[h]
	\centering
	\caption{Parameter values used in the numerical simulations.}
  \label{tab:ParamsSimu}
	\begin{tabular}{|c c c |} 
		\hline
		Parameter          & Value & Unit   \\ %[0.8ex] 
		\hline
    $\varepsilon$      & 0     & $\mu$eV \\
    $t_c$              & 15.4  & $\mu$eV \\
		$B_z$              & 24    & $\mu$eV \\
		$B_x$              & 1.62  & $\mu$eV \\
		$g_c/(2\pi)$       & 40    & MHz    \\
		$\gamma_c/(2\pi)$  & 100   & MHz    \\
		$\omega_c/(2\pi)$  & 5.85  & GHz    \\
		$\kappa/(2\pi)$    & 1.77  & MHz    \\
		$\Delta_0/(2\pi)$  & 5     & MHz    \\ [1ex]
		\hline
	\end{tabular}
\end{table}

Numerical results of qubit state initialization using the feedback approach described above will now be presented and compared to two other initialization approaches, namely thermal relaxation to the ground state and conditional feedback based on a single shot measurement.
These two other approaches are considered as they can be performed with the DQD-$\mu$WSCc architecture without any additional hardware such as a single electron transistor (SET) to measure the qubit. All numerical simulations have been performed using QuTip~\cite{QuTip2_2013}, and the parameters values used are provided in Table~\ref{tab:ParamsSimu}. These values were taken from the literature~\cite{benito2017input,mi2018coherent} and are experimentally realistic.
From Table~\ref{tab:ParamsSimu}, the following values are computed (see Appendix~\ref{app:DetailedModel}, Eqs.~\eqref{eq:Detuning}, \eqref{eq:eta}-\eqref{eq:gammas}, \eqref{eq:omegacprime}-\eqref{eq:gs}): % J'ai écrit un petit programme Python pour faire ces calculs (v. CalculsParamsAPartirTableau1_YBL.py dans le répertoire du fichier maître .tex du présent article)
$\omega_R/(2\pi) = 5.855$~GHz,
$\eta = 0.0156$, $\alpha = 78.46^\circ$, $\kappa'/(2\pi) = 1.895$~MHz, and $\gamma_s/(2\pi) = 1.501$~MHz, $\omega_c'/(2\pi) = 6.02$~MHz, $\omega_q/(2\pi) = -0.146$~GHz, and $g_s/(2\pi) = 4.89$~MHz. % On remarque que le couplage $g_s$ est plus grand que les taux de décohérence effectifs $\kappa'$ et $\gamma_s$, ce qui permet des oscillations de Rabi; le couplage n'est pas si fort que ça, mais néanmoins assez fort pour avoir des oscillations de Rabi.
The effective cavity and qubit frequencies $\omega_c'$ and $\omega_q$ are given with respect to the frame rotating at the cavity driving frequency $\omega_R$. In the laboratory frame, $\omega_R$ needs to be added to these frequencies.

\subsection{Thermal relaxation to the ground state}

In thermal relaxation to the ground state, the qubit is assumed to be in an \textit{a priori} unknown initial state and one waits sufficiently long
for it to fall into its ground state~\cite{Hanson_etal_IEDM2004,hanson2007publisher}.
%for its probability to be in the ground state to be sufficiently high
%for the probability of the qubit to be in the ground state be sufficiently high
More precisely, once the probability of measuring the qubit in the ground state is sufficiently high, the qubit is considered initialized and can be used for a computation.

\begin{figure}[ht]
	%\hfill
	\subfigure{\includegraphics[width=4.2cm]{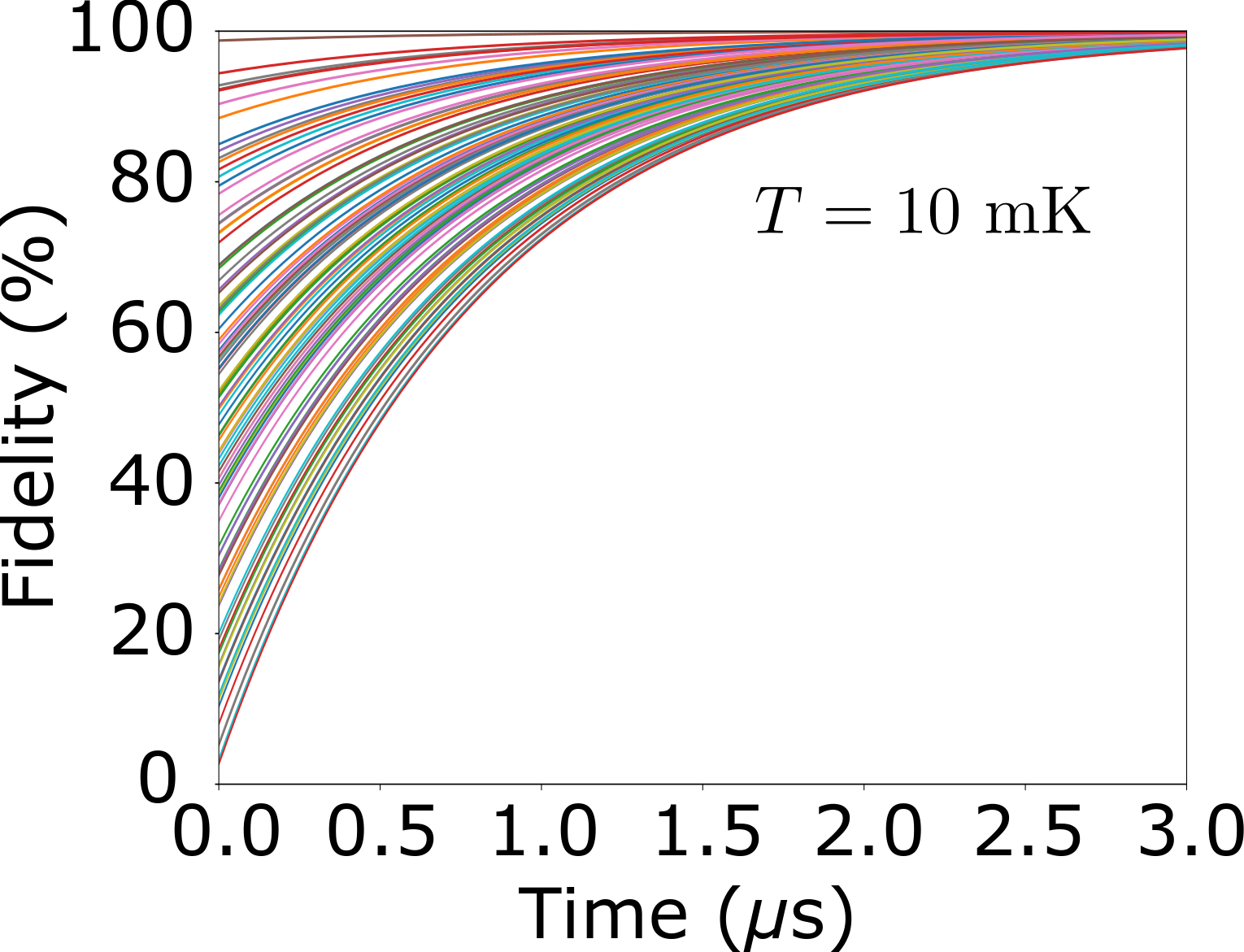}}
	%\hfill
	\subfigure{\includegraphics[width=4.2cm]{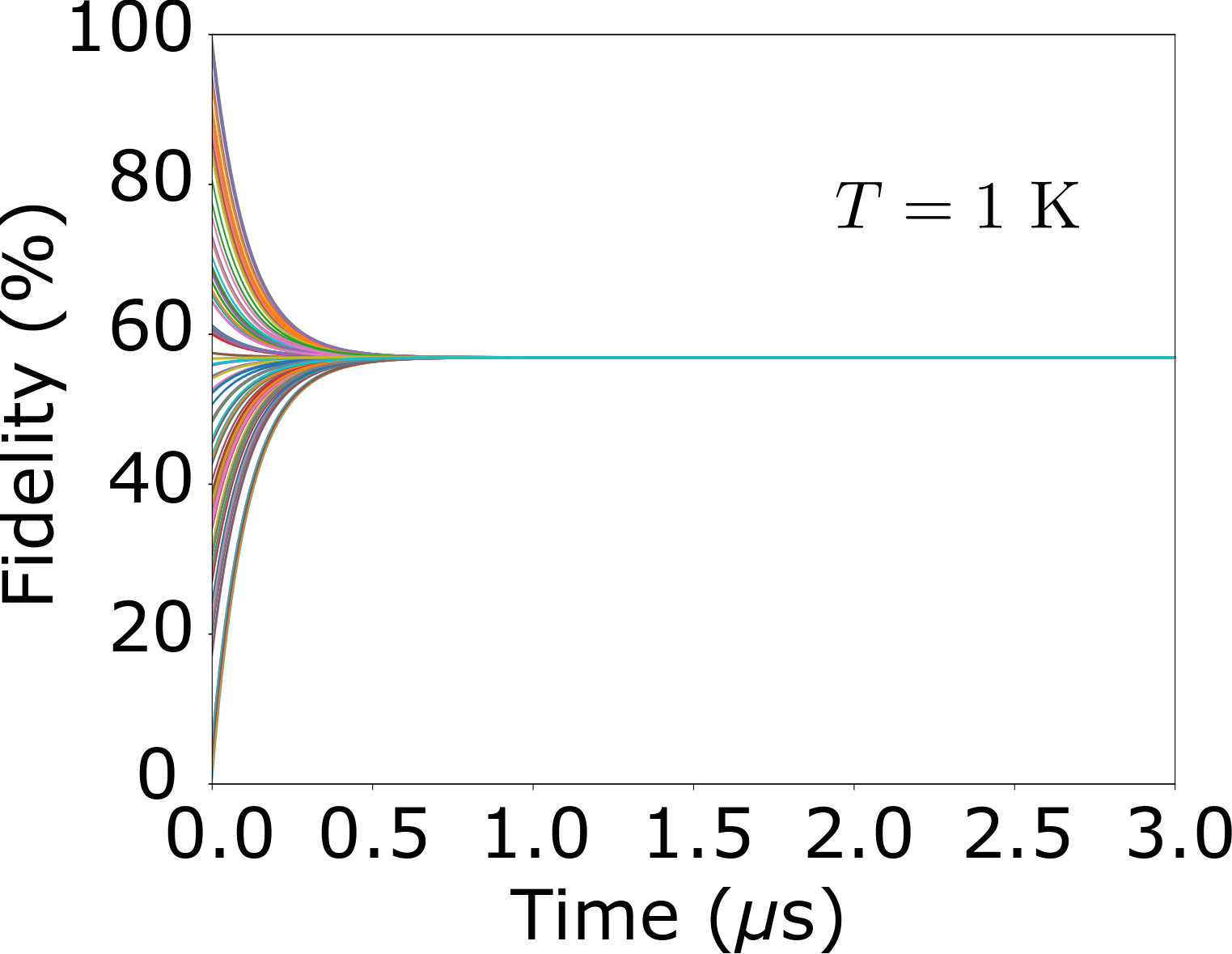}}
	%\hfill
	\caption{Qubit state initialization with thermal relaxation to the ground state.}
	\label{fig:relaxation}
\end{figure}

To model thermal relaxation along with decoherence of the qubit's state $\rho^q$, the following Lindblad equation is used
\begin{equation}
  \begin{split}
    \dot{\rho^q}
    &= -\frac{i\omega_q}{2}[\sigma_z, \rho^q] \\
    &\relphantom{=} {} -\frac{\gamma_s(1 + n_\mathrm{th})}{2}
       [\sigma_+ \sigma_- \rho^q + \rho^q \sigma_+ \sigma_- - 2\sigma_- \rho^q \sigma_+] \\
    &\relphantom{=} {} -\frac{\gamma_s n_\mathrm{th}}{2}
       [\sigma_- \sigma_+ \rho^q + \rho^q \sigma_- \sigma_+ -2\sigma_+ \rho^q \sigma_-] ,
  \end{split}
\end{equation}
from which the probability of the qubit being in the ground state at different times can be computed.
%Fig.~\ref{fig:relaxation}  depicts the fidelity graph using the relaxation method for two temperatures 10~mK and 1~K. For 10~mK it is possible to prepare the qubit in state $\ket{0} $ using the relaxation of the qubit in a time equal to 3~$\mu$s.
Fig.~\ref{fig:relaxation} shows the results at 10~mK and 1~K for 100 initial states (pure and mixed) which were randomly sampled on and within the Bloch sphere to simulate an unknown initial state.
These are realistic temperatures in experiments.
%The choice of these two temperatures corresponds to those used in the experimental conditions to manipulate a qubit.
%The initial states are randomly sampled over the surface of the Bloch sphere (pure states) and within the Bloch sphere (mixed states).
At
the
very low temperature (10~mK),
the qubit is initialized to the ground state with high probability after 3~$\mu$s irrespective of the initial state. A mean fidelity of 98.7\% is obtained with a 95\% confidence interval of [98.58, 98.82].
At the higher temperature (1~K), the approach fails reaching state $\ket{0}$ (mean initialization fidelity of 56.92\% with a 95\% confidence interval of [56.91, 56.93]).

This approach has two drawbacks.
First, it requires very low temperatures to avoid thermal excitation once in the ground state.
Second, even at very low temperatures, this approach is slow as it is limited by the longitudinal relaxation time $T_1$; optimizing this approach contradicts the requirement for qubits to have long coherence times because faster initialization requires a smaller $T_1$.

\subsection{Conditional feedback based on a single shot measurement}

\begin{figure}[ht]
	%\hfill
	\subfigure{\includegraphics[width=4.2cm]{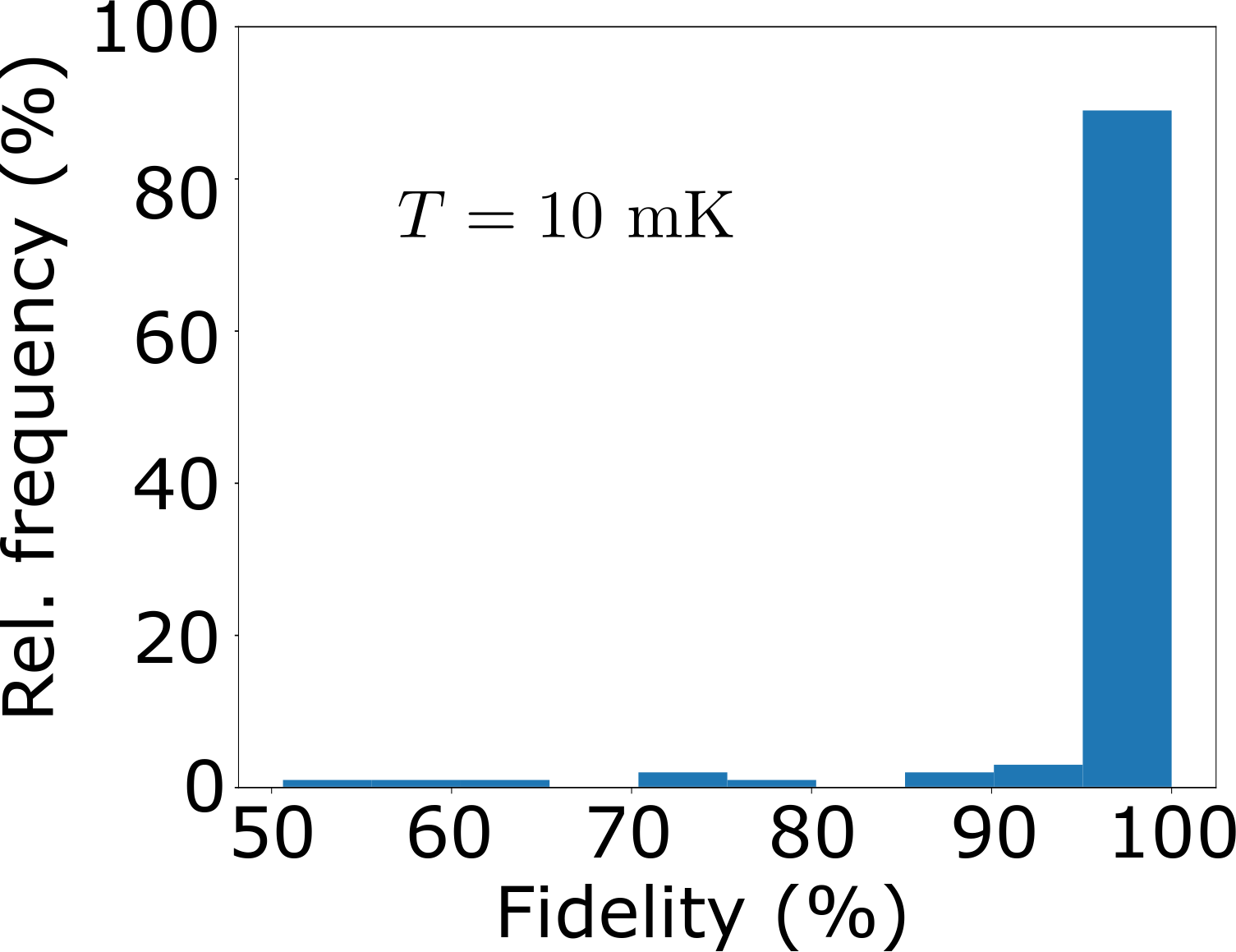}} % YBL: J'ai mis une barre de soulignement pour remplacer l'espace dans le nom du fichier.
	%\hfill
	\subfigure{\includegraphics[width=4.2cm]{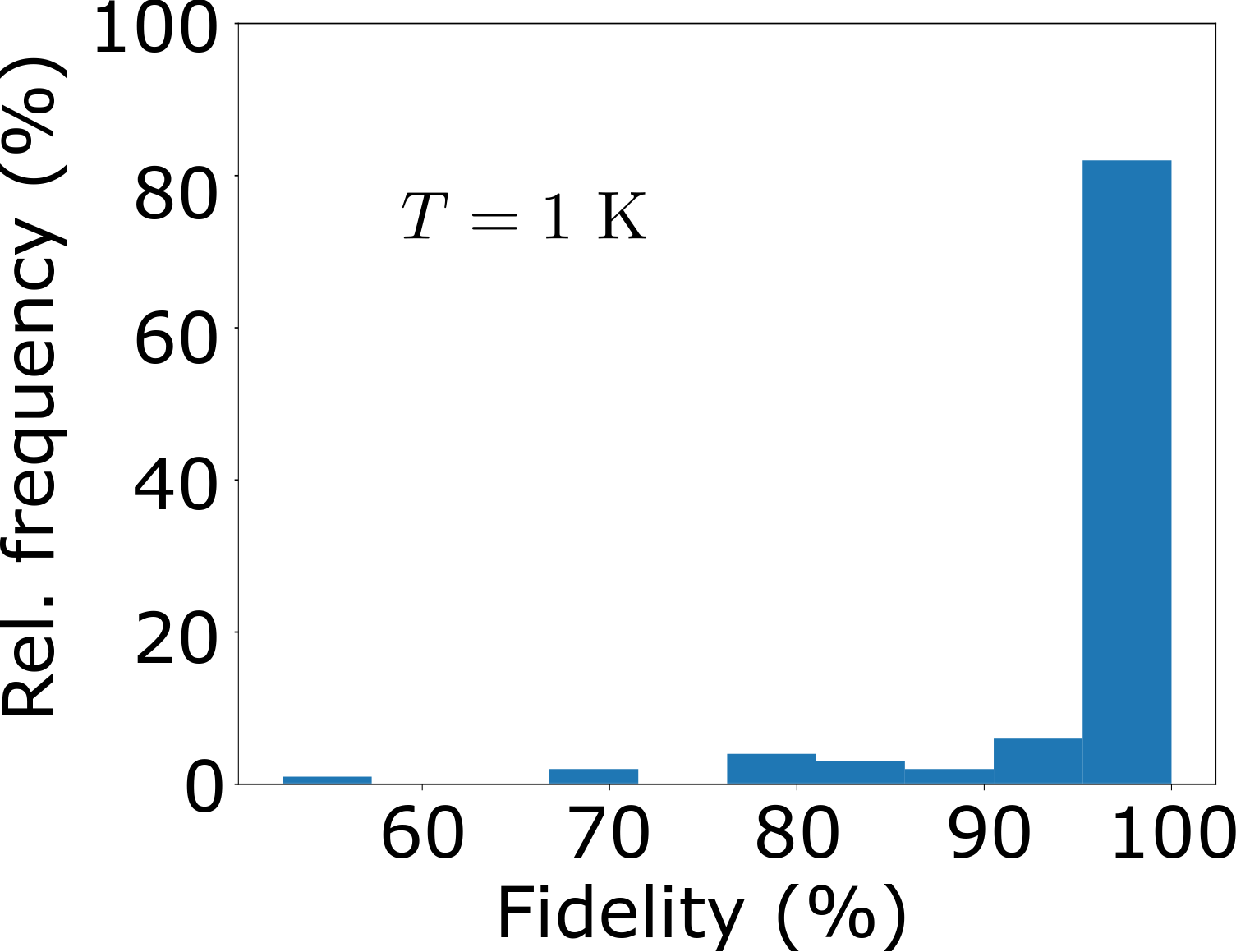}}
	%\hfill
	\caption{Qubit state initialization with conditional feedback based on a single-shot measurement.}
	\label{fig:singleshot}
\end{figure}

% On a utilisé les mêmes opérateurs de mesure que pour MBQFB, mais avec $T_m = 2~\mu$s et on calcule l'action retour sur la mesure.
Conditional feedback based on a single shot measurement
% This approach
consists of performing a QND projective, hence strong, measurement of the qubit and conditionnally applying a $\pi$ rotation (assumed to be perfect) depending on the measurement outcome~\cite{riste2012initialization,Johnson-Siddiqi_etal_HeraldStatePrep_2012,Andersen-Molmer_ClosingQFB_PRA_2016}.
If the qubit is measured in the ground state, then no action is taken and by the projection postulate the qubit is in the ground state. Else, if the qubit is found in the excited state, then a $\pi$ rotation is applied to bring the qubit to its ground state. % \textcolor{red}{(Fig.~\ref)}. J'AI PENSÉ METTRE UN SCHÉMA BLOC DE L'APPROCHE, MAIS FINALEMENT, JE NE L'AI PAS FAIT CAR ÇA PREND DE LA PLACE ET FAUT DE TEMPS
%This method consists of using a strong measurement on the qubit, if the result is state~$\ket{g}$, no action will be  performed on the qubit. If the result is state~$\ket{e}$, a rotation of angle $\pi$ is performed on the qubit.
Here a strong measurement is performed via the cavity with a measurement time $T_m = 2~\mu$s (Gaussian distributions well separated, Sect.~\ref{sect:Meas}). The same approach as in Sect.~\ref{sect:InitQFB} using the Lindblad equation is used for simulating decoherence and thermal effects during the measurement time $T_m$.
Fig.~\ref{fig:singleshot} depicts the histograms obtained with conditional feedback.
A mean fidelity of 97.14\% with 95\% confidence interval of [95.45, 98.82] is obtained at $T = 10$~mK, and at $T = 1$~K the mean fidelity is 96.50\% with a 95\% confidence interval of [95.04, 98.14].
Simulations were also performed for a measurement time $T_m = 3~\mu$s with a mean fidelity of 98.62\% (95\% confidence interval of [97.47, 99.77]) at 10~mK and mean fidelity of 98.51\% (95\% confidence interval of [97.47, 99.77]) at 1~K. Further extending the measurement time marginally improves fidelity.

%\begin{figure}[h]
%	\centering
%	\includegraphics[width=1\linewidth]{fidelity}  
%	\caption{Fidelity as function to iterations.}
%	\label{fig:fidelity}
%\end{figure}
\begin{figure}[ht]
	%\hfill
	{\includegraphics[width=4.2cm]{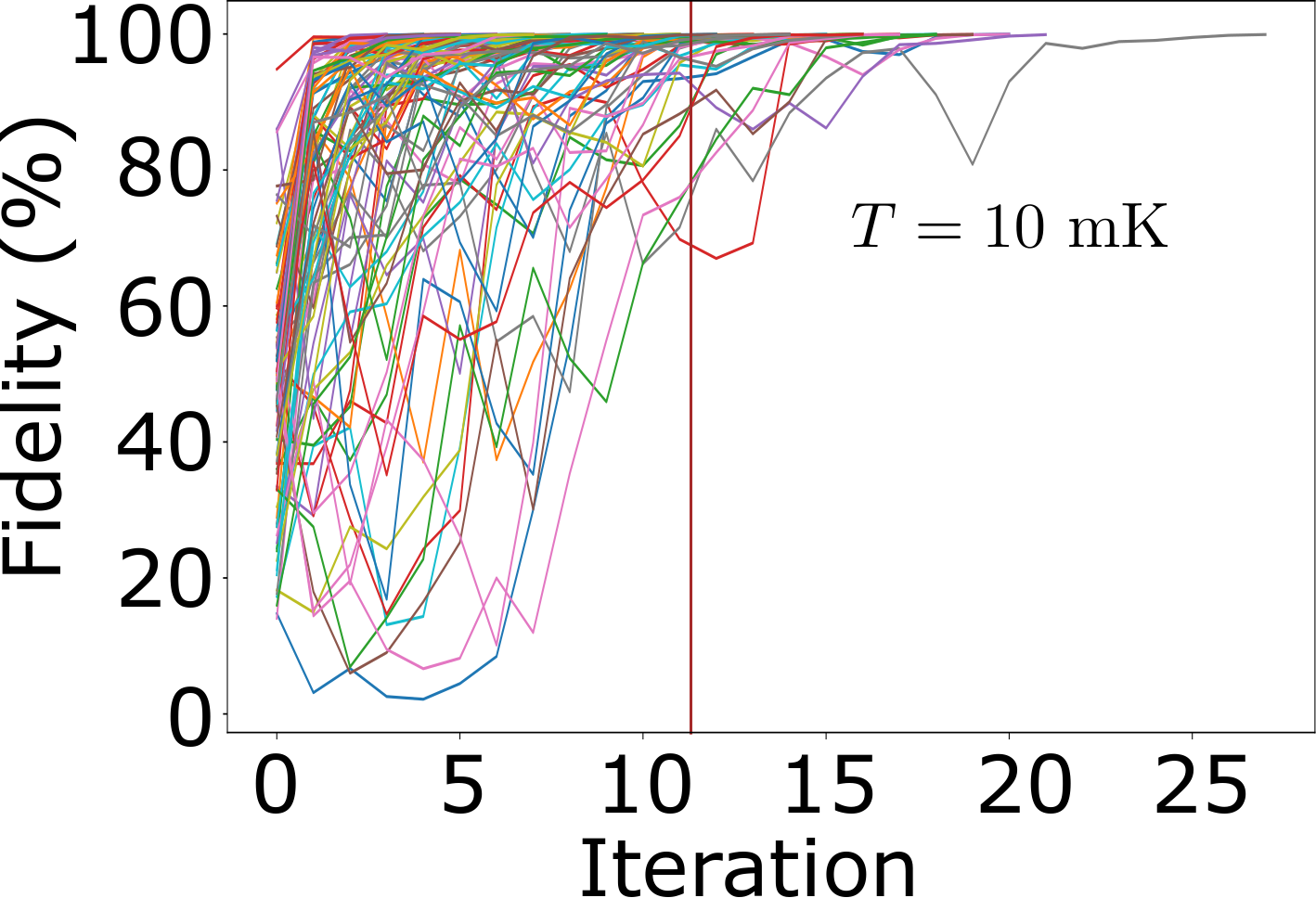}}
	%\hfill
	{\includegraphics[width=4.2cm]{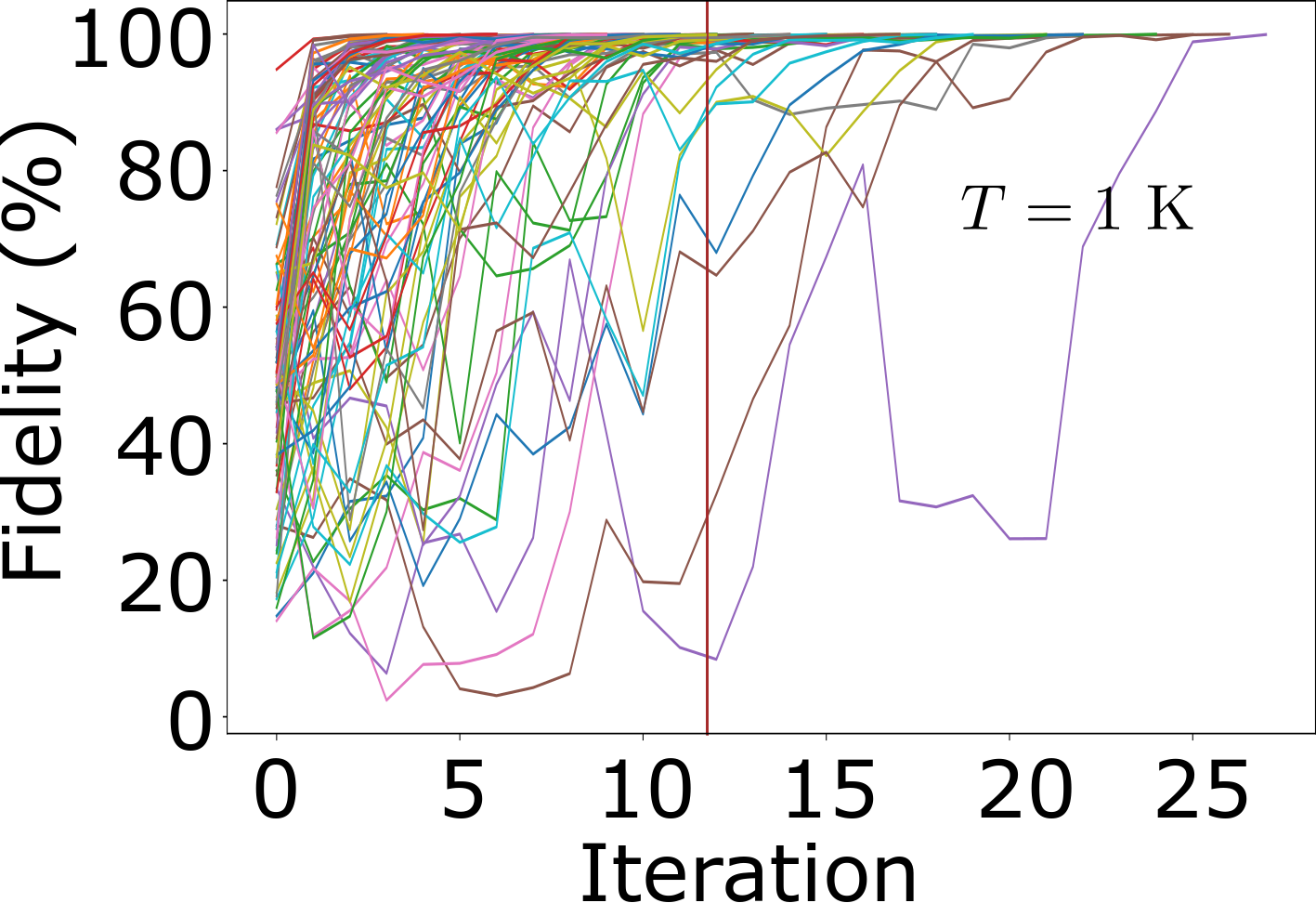}}
	%\hfill
	\caption{Fidelity as a function of feedback iterations.}
	\label{fig:fidelity}
\end{figure}

\subsection{Quantum feedback control}

In initialization relying on MBQFB control
 (Sect.~\ref{sect:InitQFB}),
%as described in (Sect.~\ref{sect:InitQFB}),
a targeted fidelity can be fixed at the outset, which is a key advantage of this approach. Here, a fidelity of 99.9\% is targeted. A weak measurement with $T_m = 200$~ns is performed in each feedback loop.
%Fig.~\ref{fig:fidelity} shows the numerical simulation results with the fidelity as a function of the number of feedback iterations.
Fig.~\ref{fig:fidelity} shows the results of the numerical simulations with the fidelity plotted as a function of the number of feedback iterations. For both temperatures (10~mK and 1~K), the mean number of iterations required to reach the target fidelity is~11 (95\% confidence interval of [10, 12]), corresponding to a mean initialization time of $2.2~\mu$s. This is shorter than the other two approaches considered above and, this, for a higher pre-set fidelity.
The figure also illustrates that the feedback state initialization protocol is effective even in the presence of quantum jumps.

\section{Conclusions}
\label{sect:Conclu}

In this work, a measurement-based quantum feedback protocol was developed for spin state initialization in an architecture relying on a gate-defined double quantum dot coupled to a superconducting cavity. This is a promising qubit building block for a spin-based quantum computer. Compared to two other initialization approaches relevant to this architecture, the protocol developed herein improves qubit state initialization as it is able to robustly initialize the spin in shorter time and reach a higher fidelity, which can be pre-set. The latter point is a highly desired feature as it provides for a more deterministic way of qubit initialization at a desired fidelity.
It is to be noted that the coupling strength $g_s$ considered here is only modestly greater than the decoherence rates $\gamma_s$ and $\kappa'$ in current experimental realizations of the DQD-$\mu$WSCc (see Table~\ref{tab:ParamsSimu} and the values computed therefrom and Fig.~4d in Ref.~\cite{mi2018coherent}). Despite this, the approach developed herein provides for fast high fidelity initialization. It can be expected that with future improvements of the couping strength, our initialization approach could even become faster.
Furthermore, the protocol developed herein is effective at high temperatures, which is critical in current efforts to scale-up the number of qubits in quantum computers.
%In addition, it promises to help solve the problem of decoherence by stabilizing qubit states.

As mentioned in the introduction, the feedback control protocol implemented here was motivated by that developed by Haroche’s group to control the state of a quantum cavity using measurements made on Rydberg atoms in the two-level approximation (akin to qubits)~\cite{dotsenko2009QFB,sayrin2011RTQFBStabPhotNbStates,guerlin2007ProgFieldStateCollapse}.
In Haroche's work, the cavity is controlled by classical microwave pulses applied to it and the state of the atoms (qubits) is measured by ionization detectors.
Here the goal was to control a qubit using measurements of a cavity. The problem tackled here is thus in a way dual to that considered by Haroche~\textit{et al.}. There are, however, fundamental differences in the measurements (necessitating the measurement theory presented in Sect.~\ref{sect:Meas}), and in the control of the qubit which is done by EDSR as presented at the beginning of Sect.~\ref{sect:InitQFB}.
The work presented here is to be best of the authors' knowledge, the first in which measurement-based quantum feedback control is applied to a spin qubit for initialization.

\begin{acknowledgements}
This project was supported by Institut quantique (IQ) at Université de Sherbrooke (UdS) through the Canada First Research Excellence Fund. RA acknowledges financial support from a Prized Postdoctoral Fellowship from IQ. VR acknowledges financial support via B1X - Bourses de maîtrise en recherche from the Fonds de recherche du Québec - Nature et technologies and Bourse VoiceAge
pour l’excellence académique aux études supérieures from UdS.
\end{acknowledgements}

\appendix

%\section*{Appendix A: Detailed DQD-$\mu$WSCc Model} % Pour le mémoire
\section{Detailed DQD-$\mu$WSCc Model}
\label{app:DetailedModel}

To understand some of the parameters used in the simulations, and for the sake of readability and completeness, this appendix reviews the mathematical/physical model of the DQD-$\mu$WSCc and the derivation that allows to arrive at the Jaynes-Cummings Hamiltonian given in Eq.~\eqref{eq:Jaynes-Cummings}. The derivations to follow are a summary, as pertinent for the present work, of those presented in Ref.~\cite{benito2017input} with some slight differences and notational changes, notably for the energy eigenstates.
The Hamiltonian for the DQD of Fig.~\ref{fig:model} is given by~\cite{benito2017input}:
\begin{equation}
H_0 = \frac{1}{2}(\varepsilon\tau_z + 2t_c\tau_x + B_z\sigma_z + B_x\sigma_x\tau_z) .
\label{eq:H0}
\end{equation}
This is a four-level system, where the $\tau$ and $\sigma$ operators are respectively Pauli operators pertaining to the electron's spatial and spin degrees of freedom.
Since the spatial degree of freedom is related to the charge's position in the double well, it is also referred to as the charge degree of freedom.
A qubit can be defined using a two-level subspace of this system.
In occupation number formalism, the $L$ and $R$ basis spatial states are defined as $\ket{L} = \ket{1,0}$ and $\ket{R} = \ket{0,1}$, where the number appearing left (right) in the ket is the occupation of the DQD's left (right) site. The spatial operators are given by: $\tau_{z}=\ket{L}\bra{L}-\ket{R}\bra{R}$, and  
$\tau_{x}=\ket{L}\bra{R}+\ket{R}\bra{L}$.

The energy levels and eigenstates of the Hamiltonian will be needed in what follows. For $B_x=0 $, there is no spin-charge coupling, and the energy orbitals are given by:
\begin{equation}
  \ket{E_0^{(0)}} = \ket{-, \downarrow},
  \ket{E_1^{(0)}} = \ket{+, \downarrow},
  \ket{E_2^{(0)}} = \ket{-, \uparrow},
  \ket{E_3^{(0)}} = \ket{+, \uparrow}.
\end{equation}
Here, superscript~(0) indicates that these are the eigenstates for $B_x = 0$, and the states $\ket{\pm}$ are the eigenstates of the bare DQD Hamiltonian
$H_\mathrm{DQD,bare} = \frac{1}{2}(\varepsilon\tau_{z}+2t_{c}\tau_{x})$ with respective eigenvalues $E_\pm = \pm \frac{1}{2}\Omega$, where $\Omega = \sqrt{\varepsilon^2 + 4t_c^2}$ is the so-called "orbital energy". Explicitly, these eigenstates are given by
\begin{align}
  \ket{-}
  &= 
  \frac{
    -\big( \cos\frac{\theta}{2} - \sin\frac{\theta}{2} \big) \ket{L}
    +
     \big( \cos\frac{\theta}{2} + \sin\frac{\theta}{2} \big) \ket{R}
  }{\sqrt{2}} ,
  \\
  \ket{+}
  &= 
  \frac{
    \big( \cos\frac{\theta}{2} + \sin\frac{\theta}{2} \big) \ket{L}
    +
    \big( \cos\frac{\theta}{2} - \sin\frac{\theta}{2} \big) \ket{R}
  }{\sqrt{2}} ,
\end{align}
where $\theta = \arctan{\frac{\varepsilon}{2t_c}} $ is the so-called "orbital angle".

For $B_x \ne 0$, the energy levels can be obtained exactly analytically and are as follows in ascending order (the calculations pose no difficulty, but are lengthy)~\cite{benito2017input}:
\begin{eqnarray}
 E_0 &=& -\frac{1}{2} \biggr[ \biggr( \Omega + \sqrt{B_z^2 + B_x^2\sin^2\theta} \,\, \biggr)^2 + B_x^2\cos^2\theta \biggr]^\frac{1}{2}, \\
 E_1 &=& -\frac{1}{2} \biggr[ \biggr( \Omega - \sqrt{B_z^2 + B_x^2\sin^2\theta} \,\, \biggr)^2 + B_x^2\cos^2\theta \biggr]^\frac{1}{2}, \\
 E_2 &=&  \frac{1}{2} \biggr[ \biggr( \Omega - \sqrt{B_z^2 + B_x^2\sin^2\theta} \,\, \biggr)^2 + B_x^2\cos^2\theta \biggr]^\frac{1}{2}, \\
 E_3 &=&  \frac{1}{2} \biggr[ \biggr( \Omega + \sqrt{B_z^2 + B_x^2\sin^2\theta} \,\, \biggr)^2 + B_x^2\cos^2\theta \biggr]^\frac{1}{2} .
\end{eqnarray} 
Analytical expressions for the associated energy eigenstates $\ket{E_0}$, $\ket{E_1}$, $\ket{E_2}$, and $\ket{E_3}$ are much more difficult to obtain analytically. Rather, first order stationary perturbation theory is resorted to; this is justified for the present purposes since $B_x \ll B_z$. To apply perturbation theory, the Hamiltonian is written as
\begin{equation}
 H_0 = H_z + H_x ,
\end{equation} 
where $H_z = H_0(B_x = 0)$ is the Hamiltonian with $B_x = 0$, and $H_x = \frac{1}{2} B_x \tau_x \sigma_z$ is the perturbation caused by the magnetic field $B_x$ and responsible for spin-charge hybridization. The $H_x $ term can be considered small compared to the elements of the $ H_z $ matrix, since $B_x \ll B_z$. To first order, the eigenstates are then given by:
\begin{equation}
  \ket{E_n^{(1)}}
  =
  \ket{E_n^{(0)}}
  +
  \sum_{k \ne n} \ket{E_k^{(0)}}
                 \frac{ \bra{E_k^{(0)}} H_x \ket{E_n^{(0)}} }{E_n^{(0)} - E_k^{(0)}},
\end{equation}
where $\ket{E_k^{(0)}}$ and $\ket{E_n^{(0)}}$ are the eigenstates of $H_z$. This leads to the following expressions for the energy eigenstates to first order (not normalized):
\begin{eqnarray}
 \ket{E_0^{(1)}}&=&\ket{-,\downarrow}+\frac{B_x\sin\theta}{2B_z}\ket{-, \uparrow}+\frac{B_x\cos\theta}{2(B_z+\Omega)}\ket{+, \uparrow},\\
 \ket{E_1^{(1)}}&=&\ket{-,\uparrow}-\frac{B_x\sin\theta}{2B_z}\ket{-, \downarrow}+\frac{B_x\cos\theta}{2(\Omega-B_z)}\ket{+, \downarrow},\\
 \ket{E_2^{(1)}}&=&\ket{+,\downarrow}-\frac{B_x\cos\theta}{2(\Omega-B_z)}\ket{-, \uparrow}-\frac{B_x\sin\theta}{2B_z}\ket{+, \uparrow},\\
 \ket{E_3^{(1)}}&=&\ket{+,\uparrow}-\frac{B_x\cos\theta}{2(B_z+\Omega)}\ket{-, \downarrow}+\frac{B_x\sin\theta}{2(B_z+\Omega)}\ket{+, \uparrow}.
\end{eqnarray}
Since $B_x \ll B_z$, the eigenstates can be further simplified to:
\begin{eqnarray}
  \ket{E_0^{(1)}}&=&\ket{-,\downarrow}, \label{eq:E0BondAntiBondBasis} \\
  \ket{E_1^{(1)}}&=&\cos\frac{\Phi}{2}\ket{-,\uparrow}+\sin\frac{\Phi}{2}\ket{+,\downarrow}, \label{eq:E1BondAntiBondBasis} \\
 \ket{E_2^{(1)}}&=&\sin\frac{\Phi}{2}\ket{-,\uparrow}-\cos\frac{\Phi}{2}\ket{+,\downarrow}, \label{eq:E2BondAntiBondBasis} \\
 \ket{E_3^{(1)}}&=&\ket{+,\uparrow} \label{eq:E3BondAntiBondBasis},
\end{eqnarray}
where $\Phi=\arctan\frac{B_x\cos\theta}{\Omega-B_z}$ is the so-called "spin-orbit mixing angle". The latter states form the orbital basis in the presence of coupling.

To obtain a coupling between the spin and microwave photons ($\sim 10$~GHz), the gap between the energy eigenstates must be of the order of $\geq $ 40~$\mu$eV. The qubit can be defined on either the transition $E_0 \leftrightarrow E_1$ or the transition $E_0 \leftrightarrow E_2$, with $E_0$ being the ground state energy.
%Here, the transition $E_0 \leftrightarrow E_1$ is chosen to define the qubit.
Charge decoherence corresponds to the transition from the state $\ket{+,\downarrow}$ to the state $\ket{-,\downarrow}$.

To manipulate the qubit state with feedback control, it is essential to use weak
%non-destructive
QND
measurements. Such measurements can be performed with the DQD coupled to a single quantized mode of frequency $\omega_c = 2\pi f_c$ of the superconducting microwave cavity (Fig.~\ref{fig:model}).
The Hamiltonian for the cavity mode of frequency $\omega_c$ is given by
\begin{equation}
  H_c = \hbar \omega_c a^{\dagger} a ,
\label{eq:Hc}
\end{equation}
where $a$ and $a^{\dagger}$ are respectively the bosonic annihilation and creation operators for the cavity photons at frequency $\omega_c$.
The coupling of the DQD with the single mode of the cavity can be described by the following interaction term~\cite{benito2017input}:
\begin{equation}
  H_I = \hbar g_c (a + a^{\dagger})\tau_{z} .
\end{equation}
The term $a+a^{\dagger}$ is proportional to the electric field within the cavity. In the eigenbasis of $H_0(B_x \neq 0)$, the interaction term $H_I$ becomes non-diagonal and takes the following form~\cite{benito2017input}:
\begin{equation}
  H_I = \hbar g_c (a + a^\dagger)\sum_{n,m=0}^3 d_{n,m} \sigma_{nm} ,
\end{equation}
where the step operators $\sigma_{nm} = \ket{E_n} \bra{E_m}$ correspond to the transitions between the energy levels of the DQD and $d_{n,m}$ are the dipole moments associated with these transitions. This coupling leads to a spin-photon coupling via spin-charge hybridization.
The full Hamiltonian of the DQD coupled with the quantized mode of the cavity is given by
\begin{equation}
H = H_0 + H_c + H_I .
\end{equation}

The DQD coupled to the cavity is an open quantum system. The DQD qubit is to be measured via a measurement of the transmission of the cavity driven at a frequency $\omega_R$. The dynamics of this coupled system thus needs to be described. For this purpose, input-output theory is used~\cite{collett1984squeezing}, which is based on the Heisenberg picture and the quantum Langevin equations. If the cavity frequency is close to the Zeeman frequency $ (\hbar \omega_c \approx B_z)$, the transition $E_0 \leftrightarrow E_3$ is off-resonance with the Zeeman frequency, and thus level $E_3$ can be ignored.
In addition, $\sigma_{03}$ satisfies $\dot{\sigma}_{03}=-(\frac{\gamma_{c}}{2})\sigma_{03}$, where $\gamma_c$ is the charge decoherence rate;  $\sigma_{03}$ is not coupled to $\sigma_{01}$ and $\sigma_{02}$. It is shown in Ref.~\cite{benito2017input} that the dynamical evolution of the operators~$a$, $\sigma_{01}$ and~$\sigma_{02}$ in the frame rotating at the driving frequency $\omega_R$ is governed by:
%
% LES ÉQS SUIVANTES SONT LES ÉQS 25, 26 ET 27 DANS BENITO
\begin{gather} 
  \dot{a} = i\Delta_0 \, a - \frac{\kappa}{2} a + \sqrt{\kappa_1} a_{in,1} - i g_c \, (d_{01}\sigma_{01} + d_{02}\sigma_{02}), \label{eq:a3} \\
\dot{\sigma}_{01} = -i \delta_1\langle \sigma_{01}\rangle-\gamma_c \sin^2\frac{\Phi}{2}\sigma_{01} +\frac{\gamma_c}{2}\sin\Phi \sigma_{02} -i g_c d_{10} \, a , \\
\dot{\sigma}_{02}=-i\delta_2 \sigma_{02}-\gamma_{c} \cos^2\frac{\Phi}{2}{\sigma}_{02} +\frac{\gamma_c}{2}\sin\Phi\sigma_{01}\ -i g_c d_{20} \, a .
\end{gather}
Here,
\begin{equation}
  \Delta_0 = \omega_R - \omega_c ,
\label{eq:Detuning}
\end{equation}
is the detuning of the driving field relative to the cavity frequency, $\delta_n = (E_n - E_0)/\hbar - \omega_R$, with $\omega_R$ being near-resonant to the $E_0 \leftrightarrow E_1$ transition; $\kappa$ is the total cavity decay rate, $\kappa_1 = \kappa/2$ is the decay rate through the input port, and $a_{in,1}$ the incoming field into the cavity.

In what follows, it will be useful to reduce the spin and charge degrees of freedom to an effective two-level system. For this purpose, it is convenient to consider the orbital basis $\ket{\pm} \otimes \ket{\uparrow\downarrow}$, and introduce the operators 
$\sigma_\tau = \ket{-,\downarrow}\bra{+,\downarrow}$ and $\sigma_s =\ket{-,\downarrow}\bra{-,\uparrow}$~\cite{benito2017input}.
Here, indices $\tau$ and $s$ respectively refer to the charge and spin degrees of freedom. It is seen that $\sigma_\tau$ is the charge flip operator and will be related to charge decoherence effects in the regime of operation considered herein, and $\sigma_s$ is the spin-flip operator with the charge remaining in the $\ket{-}$ state. Through Eqs.~\eqref{eq:E0BondAntiBondBasis}-\eqref{eq:E3BondAntiBondBasis}, $\sigma_s$ and $\sigma_\tau$ are related to $\sigma_{01}$ and $\sigma_{02}$ as follows~\cite{benito2017input}:
\begin{align}
  \sigma_{01} &\approx \cos\frac{\Phi}{2} \sigma_s + \sin\frac{\Phi}{2} \sigma_\tau , \\
  \sigma_{02} &\approx \sin\frac{\Phi}{2} \sigma_s - \cos\frac{\Phi}{2} \sigma_\tau .
\end{align}
The evolution of the system can thus be described in the orbital basis by the following equations~\cite{benito2017input}:
\begin{gather}
\dot{a}= i\Delta_{0} \, a -\frac{\kappa}{2} a + \sqrt{\kappa_1} a_{in,1} + i g_c \cos\theta \, \sigma_\tau, \label{eq:a_5} \\
\dot{\sigma}_\tau = -i \Delta_\tau \, \sigma_\tau - \gamma_c \, \sigma_\tau + i g_c \cos\theta \, a + \frac{i}{2} \frac{B_x}{\hbar}\cos\theta \, \sigma_s , \label{eq:sigmatau}
\end{gather}
\begin{gather}
\dot{\sigma}_s = -i \Delta_s \, \sigma_s +  \frac{i}{2} \frac{B_x}{\hbar}\cos\theta \, \sigma_\tau , \label{eq:sigmas}
\end{gather}
with
\begin{equation}
  \Delta_{\tau(s)} = +(-) \frac{(\Omega-B_z)/\hbar}{2} - E_0/\hbar - \omega_R ,
\end{equation}
where $\Delta_\tau/(2\pi)$ and $\Delta_s/(2\pi)$ are in the few GHz range.

To reduce the charge decoherence (\textit{i.e.} maintain superpositions of $\ket{L}$ and $\ket{R}$ states) and maximize the spin-photon coupling, it is important that the dynamics of the charge be in a stationary mode ($\dot{\sigma}_\tau = 0$, frozen charge dynamics), which gives~\cite{benito2017input}: 
\begin{equation}
\begin{split}
\dot{a}= i(\Delta_0 + \Delta_\tau \eta \cos^2\alpha) \, a - \frac{\kappa^\prime}{2} a + \sqrt{\kappa_1} a_{in,1} \\ + i \sin\alpha\cos\alpha \, \eta \, (\Delta_\tau + i\gamma_c) \, \sigma_s,
\end{split}
%\dot{a}= i(\Delta_{0}+\Delta_\tau \eta\cos^2(\alpha))a -\frac{\kappa\prime a}{2}+\sqrt{\kappa_1} b_{in,1}+i\eta\sin\alpha\cos\alpha(\Delta_{\tau}+i\gamma_c)\sigma_s,
\label{eq:QLEaOrbBasis}
\end{equation}
\begin{equation}
\dot{\sigma}_s = -i(\Delta_s - \Delta_\tau \eta \sin^2\alpha) \, \sigma_s - \gamma_s \, \sigma_s + i\sin\alpha\cos\alpha \eta \, (\Delta_{\tau}+i\gamma_c) \, a ,  
\label{eq:QLEsigmasOrbBasis}
\end{equation}
with
\begin{gather}
  \eta=\frac{(B_x/\hbar)^2 / 4 + g_c^2}{\Delta^2_{\tau}+\gamma^2_c}\cos^2\theta , \label{eq:eta} \\
  \alpha = \arctan\frac{(B_x/\hbar)}{2 g_c} , \label{eq:alphaQubitDQD}
\end{gather}
and effective decay rates
\begin{gather}
  \kappa' = \kappa + 2\gamma_c \eta \cos^2\alpha , \label{eq:kappaPrime} \\
  \gamma_s = \gamma_c \eta \sin^2\alpha . \label{eq:gammas}
\end{gather}
Because the charge dynamics are frozen, the spin degree of freedom represented by $\sigma_s$ and the cavity degree of freedom represented by $a$ will be the only degrees of freedom considered.
Neglecting $\gamma_c$ compared to $\Delta_\tau$ ($\gamma_c/(2\pi) = 100$~MHz, see Table~\ref{tab:ParamsSimu} below, whereas $\Delta_s/(2\pi)$ is in the GHz range), Eqs.~\eqref{eq:QLEaOrbBasis} and~\eqref{eq:QLEsigmasOrbBasis} can be written as~\cite{benito2017input}
\begin{gather}
\dot{a}= i \omega'_c \, a - \frac{\kappa^\prime}{2} a + \sqrt{\kappa_1} a_{in,1} + i g_s \, \sigma_s, \label{eq:QLEaOrbBasis2} \\
\dot{\sigma}_s = -i \frac{\omega_q}{2} \, \sigma_s - \gamma_s \, \sigma_s + i g_s \, a \label{eq:QLEsigmasOrbBasis2}
\end{gather}
where
\begin{gather}
  \omega'_c = \Delta_0 + \Delta_\tau \eta \cos^2\alpha , \label{eq:omegacprime} \\
  \frac{\omega_q}{2} = \Delta_s - \Delta_\tau \eta\sin^2\alpha, \label{eq:omegaq} \\
  g_s= \sin\alpha\cos\alpha \, \eta \, \Delta_\tau . \label{eq:gs}
\end{gather}
Considering the input field mode $a_{in,1}$ and the effective decay rates $\kappa'$ and $\gamma_s$, Eqs.~\eqref{eq:QLEaOrbBasis2} and~\eqref{eq:QLEsigmasOrbBasis2} show that the dynamics of the operators~$a$ and~$\sigma_s$ are those of a two-level system coupled to a single-mode field described by an effective Jaynes-Cummings Hamiltonian given by
\begin{equation}
H_\mathrm{JC} = \hbar\omega'_c \, a^\dagger a + \frac{\hbar\omega_q}{2}\sigma_z + \hbar g_s (a^\dagger \sigma_s + a \sigma_s^\dagger) .
\label{eq:Jaynes-CummingsBis}
\end{equation}
In other words, Eqs.~\eqref{eq:QLEaOrbBasis2} and~\eqref{eq:QLEsigmasOrbBasis2} can be obtained from this effective Hamiltonian.

\bibliographystyle{apsrev4-2}

%\bibliography{biblio}  % Pour la soumission, mettre cette ligne en commentaires et copier le contenu du fichier .bbl après la présente ligne

%

\end{document}